%% file: main.tex
\documentclass[11pt,letterpaper]{article}
\usepackage[letterpaper,margin=1in]{geometry}
\usepackage{amsmath,amssymb,amsthm,mathtools,bm}
\usepackage[ruled,vlined,linesnumbered]{algorithm2e}
\usepackage{tikz}
\usepackage{xcolor}
\usetikzlibrary{arrows.meta}
\usepackage{microtype}
\usepackage{enumitem}
\usepackage[numbers,square,sort&compress]{natbib}
\usepackage{xurl}
\usepackage[colorlinks=true,linkcolor=blue,citecolor=blue,urlcolor=blue]{hyperref}
\usepackage[capitalize,nameinlink,nosort]{cleveref}
\usepackage{subcaption}
\newtheorem{thm}{Theorem}[section]
\newtheorem{lem}[thm]{Lemma}
\newtheorem{cor}[thm]{Corollary}
\newtheorem{claim}[thm]{Claim}
\theoremstyle{remark}
\newtheorem{remark}[thm]{Remark}
\Crefname{lem}{Lemma}{Lemmas}
\Crefname{cor}{Corollary}{Corollaries}
\Crefname{claim}{Claim}{Claims}
\Crefname{thm}{Theorem}{Theorems}
\Crefname{remark}{Remark}{Remarks}
\Crefname{appendix}{Appendix}{Appendices}
\Crefname{algocf}{Algorithm}{Algorithms}
\newcommand{\poly}{\operatorname{poly}}
\newcommand{\Z}{\mathbb Z}
\newcommand{\Q}{\mathbb Q}
\newcommand{\R}{\mathbb R}
\newcommand{\F}{\mathbb F}

\newcommand{\ip}[2]{\langle #1,#2\rangle}
\DeclareMathOperator{\Tr}{Tr}
\DeclareMathOperator{\dist}{dist}
\DeclareMathOperator{\spn}{span}
\allowdisplaybreaks
\SetArgSty{textup}
\SetKwInOut{Input}{Input}
\SetKwInOut{Output}{Output}

\newenvironment{restatedstatement}[1]
  {\par\medskip\noindent
   \textbf{\Cref{#1} (restated).}\ \itshape}
  {\par\medskip}

\title{Exact $\ell_p$-CVP via Euclidean SVP in Polynomial Space}
\hypersetup{pdftitle={Exact lp-CVP via Euclidean SVP in Polynomial Space},pdfauthor={Jiseung Kim, Minkyu Kim, Hyung Tae Lee}}
\author{Jiseung Kim\\Jeonbuk National University\\\texttt{jiseungkim@jbnu.ac.kr} \and Minkyu Kim \\The Affiliated Institute of ETRI\\\texttt{mkkim@nsr.re.kr} \and Hyung Tae Lee\\Chung-Ang University\\\texttt{hyungtaelee@cau.ac.kr}
}
\date{}

\begin{document}
\maketitle

\begingroup
\let\small\normalsize
\input{abstract}
\endgroup

\input{1_introduction}
\input{2_preliminaries}
\input{3_short_vector_sampler}
\input{4_euclidean_cvp}
\input{5_lp_cvp}

\input{6_sampler_proof}
\input{7_deferred_cvp_proofs}
\input{8_conclusion}

\appendix
\input{a_oracle_details}
\input{ai_methodology}

\setlength{\bibsep}{2pt plus 1pt minus 1pt}
\bibliographystyle{alphaurl}
\bibliography{references}

\end{document}

%% file: abstract.tex
\begin{abstract}
We give a randomized, rank-preserving reduction from exact $\ell_p$-CVP to exact Euclidean SVP for every integer $p\ge2$. For rational inputs of bit length $\ell$ on full-rank lattices $L\subset\mathbb R^N$, the reduction uses $(Cp)^N\poly(p,\ell)$ oracle calls and additional bit operations, where $C$ is an absolute constant. It uses $\poly(p,\ell)$ working space outside the oracle, and all oracle queries have rank $N$ and $\poly(p,\ell)$ bit complexity.

For $p=2$, the reduction transfers polynomial-space exact SVP algorithms to exact CVP with $2^{O(N)}\poly(\ell)$ time overhead. Instantiating it with the SVP algorithms of Kannan and of Feng, Gao and Liu yields randomized polynomial-space CVP algorithms running in time $N^{N/(2e)+o(N)}$ and $N^{N/(4e)+o(N)}$, respectively. For every fixed integer $p\ge3$, the same reduction yields exact $\ell_p$-CVP algorithms for arbitrary rational targets either in randomized time and space $2^{O(N)}$, or in randomized time $N^{N/(4e)+o(N)}$ and polynomial space. These algorithmic bounds are up to polynomial factors in $\ell$.

The reduction performs local search sequentially over the cosets of \(2L\), where \(L\) is the input lattice, while tracking the excess of the current objective value over the optimum. Within a coset containing a closest vector, midpoint inequalities bound the unknown correction in terms of this excess, which in turn allows us to construct many short corrections that reduce the excess bound by a factor of four. A counting argument shows that these improving corrections form a sufficiently large fraction of the lattice points in a known search region, and a sparsification-based sampler assigns sufficiently large probability to every nonzero vector in that region, using at most one SVP call per sample.
Repeating this improvement over polynomially many stages yields exact recovery. For \(p\ge3\), the same framework is extended using a diagonal weighted norm computed from the current point.

\end{abstract}

%% file: 1_introduction.tex
\section{Introduction}\label{sec:introduction}
The Shortest Vector Problem (SVP) and the Closest Vector Problem (CVP) are two fundamental computational problems on lattices. A rank-$N$ lattice $L\subset\R^N$ is the set of all integer combinations of $N$ linearly independent vectors. Given such a lattice, SVP asks for a shortest nonzero vector of $L$, while CVP asks for a lattice vector closest to a given target $t\in\R^N$. We study the exact versions of these problems, in which an optimal vector must be found.

In the Euclidean norm, SVP reduces to CVP without increasing the rank~\cite{GMSS99}. In contrast, reducing CVP to SVP is particularly interesting under space constraints.
With exponential space, exact Euclidean CVP can be solved in time $2^{N+o(N)}$~\cite{ADS15full}, while exact Euclidean SVP admits faster algorithms, most recently Hhan's algorithm with time and space $2^{N/2+o(N)}$~\cite[Theorem~1.1]{Hha26}. In polynomial space, however, a substantial gap has persisted. Kannan's enumeration algorithm solves SVP in time $N^{N/(2e)+o(N)}$, whereas the corresponding CVP bound is $N^{N/2+o(N)}$~\cite{Kan83,HS07full}; Fu~\cite{Fu16} improves only a lower-order factor in the latter bound. More recently, Feng, Gao and Liu announced a randomized polynomial-space SVP algorithm running in time $N^{N/(4e)+o(N)}$~\cite{FGL26}. Thus, the best known worst-case polynomial-space running times for SVP and CVP have differed by a factor of $N^{\Omega(N)}$.

This gap reflects a limitation of known algorithms rather than a known separation between the problems.
The fastest exact CVP algorithms rely on large collections of lattice objects, such as Voronoi-relevant vectors~\cite{MVfull} or exponentially many discrete Gaussian samples~\cite{ADS15full}. This raises a natural question: can exact CVP be solved using an SVP oracle while maintaining only polynomial working space? We answer this question affirmatively with a randomized, rank-preserving reduction from exact CVP to exact Euclidean SVP that uses polynomial working space outside the oracle.

Our approach also addresses a longstanding question about exact CVP in non-Euclidean norms. For an integer $p\ge1$, let $\|z\|_p=\left(\sum_i |z_i|^p\right)^{1/p}$,
and define $\ell_p$-CVP as the problem of finding a lattice vector minimizing $\|x-t\|_p$. For every fixed integer $p\ge3$, single-exponential algorithms for exact $\ell_p$-CVP were known when the distance from the target to the lattice is at most a constant multiple of the shortest-vector length~\cite{DPV11},~\cite[Theorem~1.9]{DV12}, but not for arbitrary targets. Micciancio and Voulgaris~\cite[Section~1.2]{MVfull} explicitly identified extending single-exponential exact CVP algorithms to other norms, even with randomization, as an open problem. Bl\"omer and Kohn~\cite[Theorem~7]{BK18} showed that the number of Voronoi-relevant vectors can be unbounded even for the $\ell_3$ norm in dimension three, illustrating a fundamental obstacle to extending the Euclidean Voronoi-cell approach. For arbitrary targets, previous algorithms achieve $N^{O(N)}$ time in polynomial space~\cite{BN11exact} or $(\log N)^{O(N)}$ time via integer programming~\cite[Theorem~5]{RR23full}. Known reductions between norms~\cite{ACKLS21,EV20} incur approximation loss and therefore do not resolve the exact problem.

Our reduction addresses both gaps within a unified framework. For $p=2$, it transfers polynomial-space exact SVP algorithms to exact CVP with only single-exponential overhead, while preserving the rank. For every fixed integer $p\ge3$, it converts Euclidean SVP algorithms into arbitrary-target exact $\ell_p$-CVP algorithms, yielding either a single-exponential time-and-space instantiation or a polynomial-space instantiation. In both cases, the reduction itself uses only polynomial working space outside the oracle.

\subsection{Our Results}\label{sec:results}

We now state the reduction precisely. For $p=2$, the input consists of a positive definite rational Gram matrix $G=B^TB$ and the coordinates $\bar t\in\Q^N$ of the target with respect to $B$, so that $t=B\bar t$; the basis $B$ itself need not be given. For $p\ge3$, we consider full-rank lattices in $\R^N$ given by nonsingular rational $N\times N$ bases, together with rational targets. Let $\ell$ denote the total bit length of the input, including the encodings of $p$ and the requested failure probability $\varepsilon$.

\begin{thm}\label[thm]{thm:main}
There exist an absolute constant $C$ and a randomized algorithm with access to an exact Euclidean SVP oracle such that, for every integer $p\ge 2$, every rational $\varepsilon\in(0,1/2)$, and every input as described above, the algorithm outputs an $\ell_p$-closest vector to the target with probability at least $1-\varepsilon$. For $p=2$, the output is represented by its coefficient vector with respect to $B$.
On every execution, it
\begin{enumerate}[label=(\roman*),nosep]
\item makes at most $(Cp)^N\poly(p,\ell)$ oracle calls and performs as many additional bit operations,
\item queries the oracle only on lattices of rank $N$ given by $\poly(p,\ell)$-bit Gram matrices, and
\item uses $\poly(p,\ell)$ space outside the oracle.
\end{enumerate}
\end{thm}

With the same asymptotic bounds and success probability, the reduction can output all closest vectors exactly once.
There are at most $2^N$ such vectors, and this is tight for $L=\mathbb Z^N$ and $t=(1/2,\ldots,1/2)$.

For $p=2$, the main consequence of \Cref{thm:main} is a transfer of polynomial-space SVP bounds to exact Euclidean CVP. Plugging in the best available polynomial-space SVP algorithms gives the following concrete bounds.

\begin{cor}\label[cor]{cor:euclid-consequences}
Exact Euclidean CVP on the rank-$N$ inputs described in \Cref{thm:main} can be solved with probability at least $1-\varepsilon$ by randomized polynomial-space algorithms with running times
\[
  N^{N/(2e)+o(N)}\poly(\ell)
  \qquad\text{and}\qquad
  N^{N/(4e)+o(N)}\poly(\ell),
\]
obtained respectively from Kannan's SVP algorithm~\cite[Theorem~2]{HS07full} and the algorithm of Feng, Gao and Liu~\cite[Theorem~4.1]{FGL26}.
\end{cor}

The underlying parameterized transfer is stated in \Cref{cor:polyspace}. Together with the reduction of~\cite{GMSS99} in the opposite direction, it also shows that the randomized polynomial-space time complexities of exact Euclidean SVP and CVP agree up to a factor $2^{O(N)}\poly(\ell)$; see \Cref{cor:equiv}.

For every fixed integer $p\ge3$, \Cref{thm:main} also yields two complementary time-space tradeoffs for full-rank lattices with arbitrary rational targets.

\begin{cor}\label[cor]{cor:lp-consequences}
For every fixed integer $p\ge3$, exact $\ell_p$-CVP on full-rank lattices given by rational bases, with arbitrary rational targets, can be solved with probability at least $1-\varepsilon$ either in randomized time and space $2^{O(N)}\poly(\ell)$, or in randomized time $N^{N/(4e)+o(N)}\poly(\ell)$ and polynomial space.
\end{cor}

The first bound uses the deterministic single-exponential SVP algorithm of Micciancio and Voulgaris~\cite{MVfull} after the rank-preserving integer-basis conversion of \Cref{lem:gram-integer}. The second uses the polynomial-space algorithm of Feng, Gao and Liu~\cite[Theorem~4.1]{FGL26}, together with \Cref{lem:gram-integer,lem:bounded-error}.

\subsection{Technical Overview}\label{sec:overview}

The reduction is a randomized local search for a closest vector. We first
explain the Euclidean case. Fix a target $t$, a closest vector $y$, and a
current lattice point $x$. If $y$ were known, the vector $y-x$ would move
$x$ to the optimum in one step. Since $y$ is unknown, the algorithm
instead searches for short lattice vectors that make substantial progress
toward the optimum.

To make such a local search possible, we need to relate progress in the
objective to geometric distance from $y$. For a lattice point $x$, define
its \emph{excess} by
$\|x-t\|^2-\|y-t\|^2$.
This quantity measures how far the objective value at $x$ is above the
optimum. The algorithm does not know the excess, because it does not know
$y$; it is used only in the analysis. Ideally, a small excess would imply
that $x$ is close to $y$, because then the unknown correction $y-x$
would necessarily be short. This implication, however, is false in
general.

\paragraph{From small excess to geometric closeness.}
For example, take $L=\Z$ and $t=1/2-\delta$ with
$0<\delta<1/4$. The target lies slightly to the left of the midpoint
between $0$ and $1$, so its closest lattice vector is $y=0$. Consider
instead the neighboring lattice point $x=1$. Its excess is
$(1/2+\delta)^2-(1/2-\delta)^2=2\delta$.
As $\delta$ tends to zero, this excess becomes arbitrarily small: the two
lattice points $0$ and $1$ are almost equally good with respect to the
objective. Nevertheless,
$|x-y|=1$
remains fixed. Thus, having a small excess does not by itself imply being geometrically
close to a closest vector, and hence does not guarantee that the desired
correction $y-x$ has small length.

This difficulty disappears once we restrict attention to a single coset
of the sublattice $2L$. Suppose that $x$ and $y$ lie in the same coset, so that
$x-y\in2L$. Then, their midpoint $(x+y)/2$ is itself a lattice point.
Since $y$ is closest to $t$, this midpoint cannot be closer to $t$ than
$y$: $\left\|(x+y)/2-t\right\|^2 \ge \|y-t\|^2$.
Expanding this inequality yields a special case of the parity-clustering inequality of Aggarwal, Dadush and Stephens-Davidowitz~\cite[Lemma~5.3]{ADS15full}:
\[
    \|x-y\|^2
    \le
    2\bigl(\|x-t\|^2-\|y-t\|^2\bigr).
\]
Therefore, within the coset of $y$, an excess of at most $\Delta$ implies
$\|x-y\|^2\le 2\Delta$.
In other words, within a coset of \(2L\), a small excess guarantees that the current point is geometrically close to a closest vector, and hence that the required correction is short. Notice that the two points $0$ and $1$ in the example
above belong to different cosets of $2\Z$, so the counterexample no
longer applies.

This provides the geometric control needed for local search, but only in a coset containing a closest vector. Rather than identifying such a coset in advance, the algorithm runs the same search independently in all $2^N$ cosets of $2L$ in $L$. In a coset containing a closest vector $y$, the bound above ensures that a small excess makes the correction $y-x$ short, and the analysis below shows that the search recovers $y$ with high probability. In other cosets, no such guarantee is needed. After processing all cosets, the algorithm compares the resulting candidates by their distance to $t$ and returns the closest one.

\paragraph{From one unknown correction to many good corrections.}

It remains to explain how the search makes progress in a coset
containing a closest vector. Suppose that $x$ lies in the same coset
of $2L$ as a closest vector $y$ and has excess at most $\Delta$, and let
$\eta=y-x$. The bound above gives $\|\eta\|^2\le 2\Delta$,
so the length of the unknown correction $\eta$ is controlled by the
current excess bound. The key observation is that the algorithm does not
need to recover $\eta$ itself.

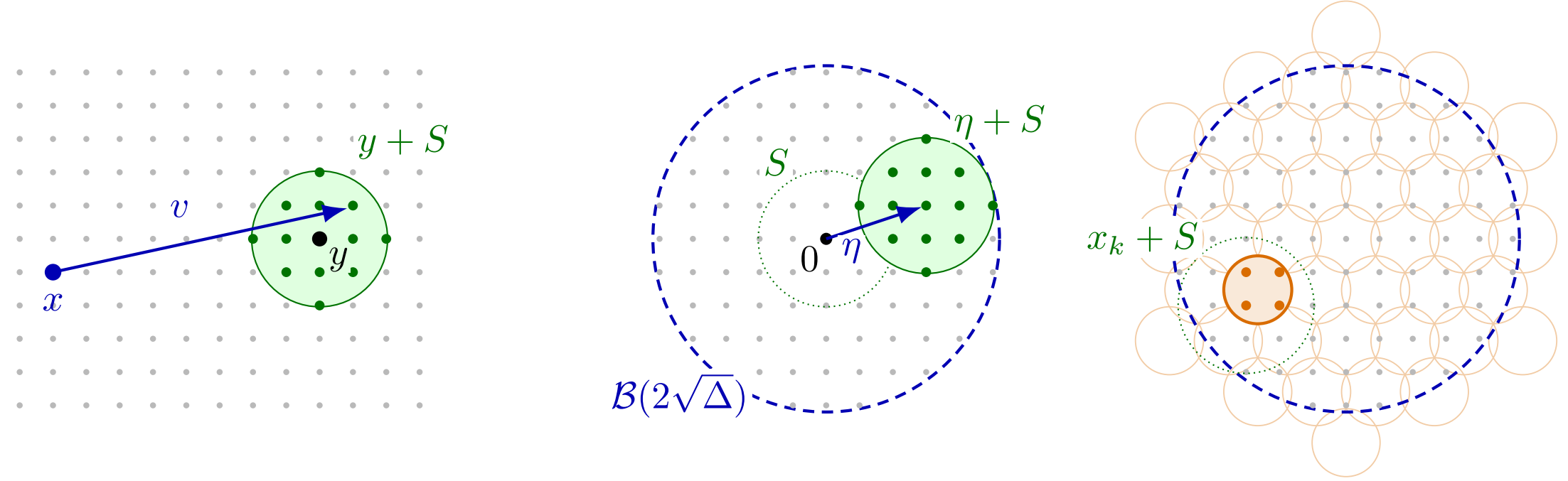
\begin{figure}[t]
\centering
\begin{tikzpicture}[x=.62cm,y=.62cm,>=Latex,font=\small]
\colorlet{cur}{blue!70!black}\colorlet{good}{green!45!black}\colorlet{cov}{orange!85!black}
\begin{scope}
  \node at (-1.6,3.95) {(a) The coset of $y$};
  \foreach \x in {-4.5,-4,...,1.5}
    \foreach \y in {-2.5,-2,...,2.5}
      \fill[gray!55] (\x,\y) circle (0.8pt);
  \fill[green!12] (0,0) circle (1.02); \draw[good] (0,0) circle (1.02);
  \foreach \x in {-.5,0,.5} \foreach \y in {-.5,0,.5} \fill[good] (\x,\y) circle (1.3pt);
  \foreach \p in {(-1,0),(1,0),(0,-1),(0,1)} \fill[good] \p circle (1.3pt);
  \fill[cur] (-4,-.5) circle (2.2pt); \node[cur,below=2pt] at (-4,-.5) {$x$};
  \fill (0,0) circle (2pt); \node[fill=green!12,inner sep=.4pt,anchor=north west] at (.1,-.1) {$y$};
  \draw[->,thick,cur] (-4,-.5)--(.42,.46); \node[cur,fill=white,inner sep=1pt] at (-2.1,.45) {$v$};
  \node[good,fill=white,inner sep=1pt] at (1.25,1.45) {$y+S$};
  \node[align=center,anchor=north] at (-1.6,-3.65) {$y+w$ has excess at most $\Delta/4$\\ for every $w\in S$};
\end{scope}
\begin{scope}[shift={(7.6,0)}]
  \node at (0,3.95) {(b) Good vectors $\eta+S$};
  \draw[cur,densely dashed,thick] (0,0) circle (2.6);
  \begin{scope}
    \clip (0,0) circle (2.62);
    \foreach \x in {-2.5,-2,...,2.5} \foreach \y in {-2.5,-2,...,2.5} \fill[gray!55] (\x,\y) circle (0.8pt);
  \end{scope}
  \draw[good,densely dotted] (0,0) circle (1.02);
  \fill[green!12] (1.5,.5) circle (1.02); \draw[good] (1.5,.5) circle (1.02);
  \begin{scope}[shift={(1.5,.5)}]
    \foreach \x in {-.5,0,.5} \foreach \y in {-.5,0,.5} \fill[good] (\x,\y) circle (1.3pt);
    \foreach \p in {(-1,0),(1,0),(0,-1),(0,1)} \fill[good] \p circle (1.3pt);
  \end{scope}
  \fill (0,0) circle (1.6pt); \node at (-.25,-.3) {$0$};
  \draw[->,thick,cur] (0,0)--(1.44,.48); \node[cur,fill=white,inner sep=1pt] at (.38,-.18) {$\eta$};
  \node[good,fill=white,inner sep=1pt] at (-.75,1.15) {$S$};
  \node[good,fill=white,inner sep=1pt] at (2.6,1.75) {$\eta+S$};
  \node[cur,fill=white,inner sep=1pt] at (-2.2,-2.35) {$\mathcal B(2\sqrt\Delta)$};
  \node[align=center,anchor=north] at (0,-3.65) {$v$ succeeds if $v\in\eta+S$,\\ and $|\eta+S|=|S|$};
\end{scope}
\begin{scope}[shift={(15.4,0)}]
  \node at (0,3.95) {(c) Counting the ball};
  \foreach \y in {0,1.530,-1.530}
    \foreach \x in {-2.650,-1.767,-.883,0,.883,1.767,2.650}
      \draw[cov!35,line width=.35pt] (\x,\y) circle (.51);
  \foreach \s in {1,-1} {
    \foreach \x in {-2.208,-1.325,-.442,.442,1.325,2.208}
      \draw[cov!35,line width=.35pt] (\x,\s*.765) circle (.51);
    \foreach \x in {-1.325,-.442,.442,1.325}
      \draw[cov!35,line width=.35pt] (\x,\s*2.295) circle (.51);
    \draw[cov!35,line width=.35pt] (0,\s*3.060) circle (.51);
  }
  \draw[cur,densely dashed,thick] (0,0) circle (2.6);
  \begin{scope}
    \clip (0,0) circle (2.62);
    \foreach \x in {-2.5,-2,...,2.5} \foreach \y in {-2.5,-2,...,2.5} \fill[gray!55] (\x,\y) circle (0.8pt);
  \end{scope}
  \fill[cov!15] (-1.325,-.765) circle (.51); \draw[cov,thick] (-1.325,-.765) circle (.51);
  \foreach \p in {(-1.5,-1),(-1.5,-.5),(-1,-1),(-1,-.5)} \fill[cov] \p circle (1.3pt);
  \draw[good,densely dotted] (-1.5,-1) circle (1.02);
  \node[good,fill=white,inner sep=1pt,anchor=south east] at (-2.15,-.3) {$x_k+S$};
  \node[align=center,anchor=north] at (0,-3.65) {the lattice points of each small\\ ball lie in some $x_k+S$, so\\ $|2L\cap\mathcal B(2\sqrt\Delta)|\le21^N|S|$};
\end{scope}
\end{tikzpicture}
\caption{Constructing and counting improving corrections in the coset of a closest vector $y$, where $\eta=y-x$.}
\label{fig:counting}
\end{figure}

Let $S=2L\cap\mathcal B(\sqrt{\Delta/6})$,
where $\mathcal B(r)$ denotes the Euclidean ball of radius $r$ centered
at the origin. For every $w\in S$, the point $y+w$ has excess at most
$\Delta/4$: by \Cref{lem:parity-geometry},
\[
  \|y+w-t\|^2-\|y-t\|^2
  \le \tfrac32\|w\|^2
  \le \tfrac{\Delta}{4}.
\]
Starting from $x$, the same point is reached by the correction $\eta+w$,
since $x+(\eta+w)=y+w$.
Hence, every vector in the translated set $\eta+S$ is a successful
correction that reduces the excess bound from $\Delta$ to $\Delta/4$.
Thus, instead of relying on the single unknown correction $\eta$, we have
an entire set $\eta+S$ of successful corrections, with
$|\eta+S|=|S|$.
\Cref{fig:counting}(a) illustrates the good points around $y$, while \Cref{fig:counting}(b) shows the corresponding corrections in $\eta+S$.

It therefore suffices to draw a random vector $V\sim\mathcal D$ from a distribution $\mathcal D$ on $2L$ such that $V\in\eta+S$ with sufficient probability; a realized sample is denoted by $v$ in \Cref{fig:counting}.
All vectors of $\eta+S$ have length less than $2\sqrt\Delta$, a radius that the algorithm knows. To compare the $|S|$ successful corrections with all lattice vectors in this search ball, cover $\mathcal B(2\sqrt\Delta)$ by balls of radius $\frac12\sqrt{\Delta/6}$. By \Cref{lem:cover}, at most $21^N$ such balls are needed. If one of these smaller balls contains a lattice point $x_k\in2L$, then any other lattice point $x\in2L$ in the same ball satisfies $\|x-x_k\|\le\sqrt{\Delta/6}$.
Since $x-x_k\in2L$, we have $x-x_k\in S$, or equivalently $x\in x_k+S$. Thus, every smaller ball contains at most $|S|$ points of $2L$ (\Cref{lem:count}), and hence
\[
 |2L\cap \mathcal B(2\sqrt\Delta)|\ \le\ 21^N\,|S|
\]
for every lattice, independently of its shape.
This counting argument is illustrated in \Cref{fig:counting}(c).
Consequently, if $\mathcal D$ assigns probability at least $\gamma/|2L\cap \mathcal B(2\sqrt\Delta)|$ to every nonzero point of $2L\cap \mathcal B(2\sqrt\Delta)$, then $\Pr[V\in\eta+S]$ is at least $\gamma\,21^{-N}/2$ (\Cref{lem:euclid-contraction}). Neither $\mathcal D$ nor the radius depends on $y$.

\paragraph{Sampling only what the reduction needs.}

We now need a way to sample vectors from the search ball using an SVP
oracle such that every nonzero lattice vector in the search ball has a
sufficiently large chance of being returned. Some vectors may be sampled
more often than others; the reduction only needs to ensure that none of
them is too unlikely. The sampler of \Cref{thm:sampler} provides exactly
this guarantee, with $\gamma=2^{-O(N)}/\poly(\ell)$, using at most one
exact SVP call per sample.

The basic idea is to use lattice sparsification to isolate one short
lattice direction. Fix any nonzero vector $z$ in the search ball.
There is a unique way to write $z=mu$,
where $m\ge1$ is an integer and $u$ is primitive, meaning that $u$ is
not itself an integer multiple of another lattice vector. Thus, to
produce $z$, it is enough to recover the direction $u$ and guess the
multiplier $m$.

To recover $u$, the sampler chooses a prime $q$ and imposes a random
linear equation modulo $q$ on the coefficient vectors of the lattice.
Only the vectors satisfying this equation are kept, and they form a
full-rank sublattice. A fixed primitive vector survives this filtering
with probability $1/q$. Thus, $q$ controls how strongly the lattice is
sparsified: increasing $q$ removes more competing vectors, but also
makes it less likely that $u$ itself survives.

For an appropriate scale of $q$, there is sufficient probability that
$\pm u$ survive while all other vectors that could be as short as $u$
are removed. On this event, $\pm u$ are the unique shortest nonzero
vectors of the sparsified lattice, so a single SVP call returns $u$ up
to sign. Multiplying the result by the guessed value of $m$ then gives
$\pm z$.

The sampler does not know the correct multiplier $m$ or the appropriate
scale of $q$ in advance, so it guesses both. 
The analysis shows that, despite these guesses, every nonzero vector in the search ball is returned with sufficiently large probability.
This sparsification step follows the
primitive-vector sampling approach of
Stephens-Davidowitz~\cite[Section~4.1]{Ste16full}, building on
Khot~\cite{Kho05} and Dadush and Kun~\cite{DK16};
see \Cref{rem:ste16} for a brief comparison and \Cref{sec:sampler-proof} for further details.

\paragraph{From local progress to exact recovery.}
The search proceeds in stages with excess bounds decreasing by a factor of four. In the coset containing $y$, $2^{O(N)}\poly(\ell)$ samples suffice at each stage to obtain the next-scale improvement with high probability. After polynomially many stages, the excess bound is below $\lambda_1(2L)^2/2$, and the midpoint inequality forces the current point to equal $y$. Since the $2^N$ cosets and all samples are processed sequentially, the reduction stores only a constant number of lattice points, one sampler workspace, and a few counters, yielding polynomial working space outside the oracle.

\paragraph{The $\ell_p$ case.}
For an integer $p\ge3$, the same strategy still applies: work within cosets of $2L$, show that many short corrections make a constant-factor improvement, and use the same sparsification sampler to hit one of them. The new difficulty is geometric. 
The Euclidean length of a correction alone no longer provides the control needed to bound its effect on $F_t(z)=\sum_i |z_i-t_i|^p$.
A displacement in a coordinate with large residual $|z_i-t_i|$ is more expensive than the same displacement in a coordinate with small residual.

We therefore replace the Euclidean norm by a diagonal weighted norm computed from the current point $x$:
\[\textstyle
  \|w\|_{H_x}^2=\sum_i h_iw_i^2,
  \qquad
  h_i=|x_i-t_i|^{p-2}+\rho,
  \qquad
  \rho=\Theta\bigl((\Delta/N)^{1-2/p}\bigr).
\]
The curvature relevant to perturbations around the optimum depends on the unknown minimizer $y$, rather than the current point $x$.
Midpoint inequalities are what make the construction usable: if $x$ and $y$ lie in the same coset and $F_t(x)-F_t(y)\le\Delta$, then they imply that $y-x$ is short in the computable norm $H_x$ and that the weights at $x$ control those around $y$ (\Cref{lem:even-midpoint,lem:weighted}).

The good perturbations around $y$ are now chosen from a coordinate box rather than a ball. This gives coordinatewise control of the Taylor remainder of $F_t$ without losing an $N^{\Omega(N)}$ factor in volume. By symmetry, at least half of the lattice points in this box also make the first-order term nonpositive. 
A covering argument then shows that a $(Cp)^{-N}$ fraction of the lattice points in a known $H_x$-ball yield the same factor-four improvement (\Cref{lem:lp-contraction}). Since \Cref{thm:sampler} works for any rational positive-definite metric, the same sparsification sampler applies to $H_x$. Finally, the rationality of the input makes the possible values of $F_t$ discrete, so once the maintained excess bound is smaller than the minimum positive difference between distinct objective values, the current point is exactly optimal.

\subsection{Related Work}\label{sec:intro-prior}

Ajtai, Kumar and Sivakumar~\cite{AKS02} solve $(1+\varepsilon)$-approximate CVP in time $2^{O(N(1+1/\varepsilon))}$ and exponential space. Nasz\'odi and Venzin~\cite{NV22} solve $(1+\varepsilon)$-approximate CVP in $\ell_p$ norms for fixed $p$, whereas our result concerns exact $\ell_p$-CVP. Arya, da Fonseca and Mount~\cite{AFM24} give a randomized $(1+\varepsilon)$-CVP algorithm for arbitrary norms with running time $2^{O(N)}\varepsilon^{-(N-1)/2}$, up to polynomial factors in the input size. Dubey and Holenstein~\cite{DH11} give a polynomial-time Turing reduction from $\gamma^2\sqrt N$-approximate Euclidean CVP to $\gamma$-approximate Euclidean SVP. 
The CVP-to-SVP reductions of~\cite{ACKLS21} incur approximation loss and increase the lattice rank from $N$ to $N+1$.
Dadush and Micciancio~\cite[Section~5.4]{DM13full} discuss a potential application of densest-sublattice preprocessing to faster polynomial-space enumeration for CVP with preprocessing, under a heuristic assumption. Hunkenschr\"oder, Reuland and Schymura~\cite{HRS20} solve CVP in polynomial space given a basis in which every facet normal of the Voronoi cell has coefficients bounded by some $c$, in time $(2c+1)^{O(N)}\poly(N)$. For Haar-random lattices, Laarhoven~\cite[Theorem~1.8]{Laa26} obtains a polynomial-space exact CVP with preprocessing (CVPP) algorithm with query time $N^{0.0892N+o(N)}$, excluding preprocessing costs.
Stephens-Davidowitz~\cite{Ste16full} gives polynomial-time, dimension-preserving reductions from centered discrete Gaussian sampling to SVP and from shifted discrete Gaussian sampling to CVP. The exact CVP algorithm of~\cite{ADS15full} uses shifted samples and requires $2^{N+o(N)}$ time and space, so these results together do not give a polynomial-space reduction from CVP to SVP.
Dadush, Eisenbrand and Rothvoss~\cite{DER25} make approximate integer programming exact when the residues of a solution modulo $\Theta(N)$ are known, and guessing these residues costs $N^{O(N)}$. 
In contrast, our reduction enumerates only the $2^N$ residue classes modulo~$2$, using midpoint inequalities specific to the $\ell_p$ objective.

\paragraph{Organization.}
\Cref{sec:setup} introduces notation and two counting lemmas; \Cref{sec:sampler-statement} presents the short-vector sampler and states its guarantees;
and \Cref{sec:euclidean,sec:even} prove \Cref{thm:main}. \Cref{sec:sampler} proves the sampling theorem, \Cref{sec:cvp-deferred} gives the deferred CVP proofs and polynomial-space consequences, and \Cref{app:implementation} treats oracle implementation details.

%% file: 2_preliminaries.tex
\section{Preliminaries}\label{sec:setup}

We introduce the notation and geometric tools used throughout the paper.

\paragraph{Lattices.}
A lattice of rank $N$ is $L=B\Z^N$ for a nonsingular $B\in\R^{N\times N}$ with columns $b_1,\ldots,b_N$. Its Gram matrix is $G=B^TB$, and we write $\lambda_1(L)$ for the minimum length of a nonzero vector of $L$.

For the Euclidean reduction, a lattice vector $Bz$ is represented by its coefficient vector $z\in\Z^N$, and the target $t=B\bar t$ by $\bar t\in\Q^N$. Thus,
$\|Bz-t\|^2=(z-\bar t)^TG(z-\bar t)$.
In our reduction for $p=2$, only the positive definite rational matrix $G$ and the rational vector $\bar t$ are given; a basis $B$ with $B^TB=G$ is never computed.
We will use a simple lower bound on the least eigenvalue of $G$ to convert length bounds into bounds on coefficient vectors. Put
$\iota(G)=\det G / (\Tr G)^{N-1}$.
Since each of the remaining $N-1$ eigenvalues is at most $\Tr G$,
$\det G\le\lambda_{\min}(G)(\Tr G)^{N-1}$,
and hence $\iota(G)\le\lambda_{\min}(G)$. Therefore,
\begin{equation}\label{eq:iota}
 z^TGz\ \ge\ \iota(G)\,\|z\|^2\qquad\text{for all }z\in\R^N.
\end{equation}
When $G$ is rational, $\iota(G)$ is computed exactly from $G$. In particular, $\lambda_1(L)^2\ge\iota(G)$, because every nonzero lattice vector has a nonzero integer coefficient vector.

We write $\dist(t,L)=\min_{x\in L}\|x-t\|$, and $\log$ for the natural logarithm. The cosets of $2L$ in $L$ are
$Ba+2L$, $a\in\{0,1\}^N$.

\paragraph{Norms given by positive definite matrices.}
For symmetric matrices $A$ and $B$, we write $A\succeq B$ if $A-B$ is positive semidefinite, and $A\succ B$ if $A-B$ is positive definite. In particular, $H\succ0$ means that $H$ is positive definite. For a symmetric positive definite matrix $H$, which we call a \emph{metric}, let $\|x\|_H=(x^THx)^{1/2}$, $\mathcal B_H(r)=\{x\in\R^N:\|x\|_H\le r\}$ and $\lambda_1(L,H)=\min_{0\ne x\in L}\|x\|_H$. We drop $H$ when $H=I$. For $L=B\Z^N$, we have $\|Ba\|_H^2=a^T(B^THB)a$, so the pair $(L,H)$ is described by the Gram matrix $B^THB$. In our algorithms, this matrix is rational and squared lengths are compared exactly.

\paragraph{SVP oracles.}
An SVP oracle receives a positive definite rational Gram matrix and returns the coefficient vector of a shortest nonzero vector. When there are several shortest vectors, the oracle may choose among them by any rule that depends on the history of the computation and on independent random bits. Our guarantees hold for every such rule. 
In \Cref{app:implementation}, \Cref{lem:gram-integer} converts a rational Gram matrix into an integer basis of the same rank whose shortest coefficient vectors are shortest for the original matrix, and \Cref{lem:bounded-error} shows how to simulate the exact oracle using an SVP algorithm that succeeds with probability $2/3$.

\paragraph{Primitive vectors.}
A nonzero $u\in L$ is \emph{primitive} if $u/k\notin L$ for every integer $k\ge2$, equivalently if the greatest common divisor of its coefficients is $1$. Every nonzero $x\in L$ equals $mu$ for exactly one integer $m\ge1$ and one primitive $u\in L$. Following~\cite{Ste16full}, for $s>0$ we write $\xi_{L,H}(s)$ for the number of primitive vectors of $L$ in $\mathcal B_H(s)$, counting each pair $\{u,-u\}$ once.

\paragraph{Counting lattice points.}
Let $\omega_N=\pi^{N/2}/\Gamma(N/2+1)$ be the volume of the Euclidean unit ball, so that $\operatorname{vol}(\mathcal B(R))=\omega_N R^N$. Since $\Gamma(z+1)\ge(z/e)^z$ for $z>0$,
\begin{equation}\label{eq:vn}
 \omega_N\ \le\ (2\pi e/N)^{N/2}.
\end{equation}
The next two lemmas are standard packing and covering bounds; their proofs are deferred to \Cref{sec:cvp-deferred}. 
They compare the numbers of lattice points in two sets of different sizes or shapes and require no assumption on the lattice. If a large set is covered by $T$ translates of half of a small symmetric convex set $Q$, then it contains at most $T$ times as many lattice points as $Q$.

\begin{lem}\label[lem]{lem:count}
Let $L\subset\R^N$ be a lattice and $Q\subset\R^N$ a convex, compact, symmetric set. If $X\subseteq\bigcup_{k=1}^T(c_k+\frac12Q)$ for some $c_1,\ldots,c_T\in\R^N$, then $|L\cap X|\le T\cdot|L\cap Q|$.
\end{lem}

\begin{lem}\label[lem]{lem:cover}
The following covering bounds hold.
\begin{enumerate}[label=(\roman*)]
\item For every positive definite metric $H$ and $R,\delta>0$, the ball $\mathcal B_H(R)$ is contained in a union of at most $(1+2R/\delta)^N$ translates of $\mathcal B_H(\delta)$.
\item For the Euclidean norm, $R>0$ and $0<\alpha\le R/\sqrt N$, the ball $\mathcal B(R)$ is contained in a union of at most $\omega_N\cdot(2R/\alpha)^N$ translates of the cube $[0,\alpha]^N$.
\end{enumerate}
\end{lem}

%% file: 3_short_vector_sampler.tex
\section{Sampling Short Lattice Vectors with an SVP Oracle}\label{sec:sampler-statement}

Throughout this section, $K$ denotes the lattice given to the sampler, to distinguish it from the CVP lattice $L$ used in \Cref{sec:euclidean,sec:even}. The progress arguments in those sections require only a pointwise lower bound on the probability of every nonzero lattice vector in a known ball. More precisely, for a distribution $\mathcal D$ on $K$, a radius $r>0$, and a number $\gamma\in(0,1]$, we require
\begin{equation}\label{eq:dom}
 \mathcal D(x)\ \ge\ \frac{\gamma}{|K\cap \mathcal B_H(r)|}\qquad\text{for every }x\in(K\cap \mathcal B_H(r))\setminus\{0\}.
\end{equation}
If $V\sim\mathcal D$, then summing over $x$ in \eqref{eq:dom} gives $\Pr[V\in E]=\mathcal D(E)\ge\gamma\,|E\setminus\{0\}|/|K\cap \mathcal B_H(r)|$ for every $E\subseteq K\cap \mathcal B_H(r)$. No upper bound on $\mathcal D(x)$ is needed, and $\mathcal D$ may assign large probability to $0$.
\Cref{alg:sampler} gives the sampler used in the reduction; the following theorem states its guarantees.

\begin{thm}\label[thm]{thm:sampler}
Let $K=B_K\Z^N$ and let $H\succ0$ be a metric such that $G_K:=B_K^THB_K\in\Q^{N\times N}$.
Let $r^2>0$ and $\lambda_0^2>0$ be rational numbers satisfying $\lambda_0\le\lambda_1(K,H)$. There is a randomized algorithm with access to an exact SVP oracle that, given $G_K$, $r^2$, and $\lambda_0^2$, outputs a coefficient vector representing a lattice vector in $K$. Let $\mathcal D$ be the induced distribution of this lattice vector, and let $V\sim\mathcal D$. Then the following properties hold.
\begin{enumerate}[label=(\alph*)]
\item\label{it:support} $\operatorname{supp}(\mathcal D)\subseteq K\cap \mathcal B_H(r)$; equivalently, $V\in K\cap \mathcal B_H(r)$ with probability $1$.
\item\label{it:dom} $\mathcal D$ satisfies~\eqref{eq:dom} with
\[
 \gamma=\frac{9^{-N}}{512\,N\,\bigl(\log_2\max\{1,r/\lambda_0\}+4\bigr)^2},
\]
conditionally on any history of earlier computations and for every rule by which the oracle chooses among shortest vectors.
\item\label{it:res} The algorithm makes at most one oracle call, on a full-rank sublattice of $K$ in the metric $H$ given by a Gram matrix whose bit length is polynomial in that of the input. Its other bit operations and its space are polynomial in the bit lengths of the input and of the oracle answer.
\end{enumerate}
\end{thm}

The reductions supply a computable lower bound $\lambda_0$ on $\lambda_1(K,H)$; the proof of \Cref{thm:sampler} is deferred to \Cref{sec:sampler}.

\paragraph{Construction.}
Write every nonzero $x\in K$ uniquely as $x=mu$ with $u$ primitive. The sampler guesses $m$ and a sparsification scale, and uses a random congruence to isolate $\pm u$ in a full-rank sublattice. Writing $K=B_K\Z^N$, for a prime $q$ and $z\in\F_q^N$ define
\[
 \Lambda_z=\{a\in\Z^N:\ip za\equiv0\pmod q\},\qquad K_z=B_K\Lambda_z .
\]
If $D_z$ is a basis matrix of \(\Lambda_z\), then the query Gram matrix is $D_z^T G_K D_z$ and an oracle answer $a'$ corresponds to the coefficient vector $D_z a'$ in $K$.

\begin{algorithm}[t]
\DontPrintSemicolon
\Input{Rational $G_K\succ0$, $r^2>0$, and $0<\lambda_0^2\le\lambda_1(K,H)^2$}
\Output{Coefficient vector of a point of $K\cap \mathcal B_H(r)$}
\vspace{2mm}
\lIf{$r^2<\lambda_0^2$}{\Return{$0$}\label{ln:trivial}}
$J_{\rm s}\gets\max\{\kappa\ge0:4^\kappa\lambda_0^2\le r^2\}$\label{ln:js}\;
$i_{\max}\gets N(J_{\rm s}+4)+1$\;
choose \(\kappa\in\{0,\ldots,J_s\}\), \(i\in\{1,\ldots,i_{\max}\}\), and then \(m\in[2^\kappa,2^{\kappa+1})\) by procedure~(1), using independent random bits\label{ln:guess}\;
choose a prime $q\in[2^i,2^{i+1})$ by procedure (2) and $z\in\F_q^N$ by procedure (3)\label{ln:sparse}\;
\lIf{any of procedures~(1)--(3) fails}
{\Return{$0$}}\label{ln:random-fail}
compute $D_z$ as defined at the beginning of Section~\ref{sec:sampler}\label{ln:compute-dz}\;
$a\gets D_za'$, where $a'$ is the oracle answer on $D_z^TG_KD_z$\label{ln:oracle}\;
\lIf{$m^2a^TG_Ka\le r^2$}{\Return{$ma$ or $-ma$, each with probability $1/2$}\label{ln:out}}
\Return{$0$}\label{ln:zero}
\caption{Sampling short lattice vectors}
\label{alg:sampler}
\end{algorithm}

For a fixed $x=mu\in K\cap\mathcal B_H(r)$, one choice of the multiplier range and prime scale makes the random congruence isolate $\pm u$ with sufficient probability. 
The basis $D_z$, the bounded-randomness procedures (1)–(3), the counting and isolation bounds, and the exact calculation of~\eqref{eq:dom} are specified in \Cref{sec:sampler}.

\begin{remark}\label[remark]{rem:ste16}
The sparsification step follows Stephens-Davidowitz~\cite[Section~4.1]{Ste16full}. Our reduction does not use the shifted/discrete-Gaussian part of that work: it needs only centered short-vector samples and the pointwise bound~\eqref{eq:dom}. The detailed comparison is in \Cref{sec:sampler}.
\end{remark}

%% file: 4_euclidean_cvp.tex
\section{Exact CVP in the Euclidean Norm}\label{sec:euclidean}

We first prove \Cref{thm:main} for $p=2$. Throughout the section, $L=B\Z^N$ has positive definite rational Gram matrix $G$, and $t=B\bar t$ with $\bar t\in\Q^N$.

\subsection{Geometry and One-Step Progress}\label{sec:euclid-geometry}

The following two lemmas formalize the Euclidean ingredients of \Cref{sec:overview}; full proofs are deferred to \Cref{sec:euclid-deferred}.
The consequence $\|x-y\|^2\le 2\Delta$ in \Cref{lem:parity-geometry} is the exact-CVP specialization of the parity-clustering inequality of Aggarwal, Dadush and Stephens-Davidowitz~\cite[Lemma~5.3]{ADS15full}; here we use it as part of the construction of a sufficiently large set of improving corrections.

\begin{lem}\label[lem]{lem:parity-geometry}
Let $y$ be a closest vector of $L$ to $t$. For every $w\in2L$,
\[
 \tfrac12\|w\|^2\ \le\ \|y+w-t\|^2-\|y-t\|^2\ \le\ \tfrac32\|w\|^2 .
\]
Consequently, for every $\Delta\ge0$ and every $x\in y+2L$ with $\|x-t\|^2\le\|y-t\|^2+\Delta$,
\begin{enumerate}[label=(\roman*),nosep]
\item\label{it:geo-close} $\|x-y\|^2\le2\Delta$;
\item\label{it:geo-equal} if $\Delta<\lambda_1(2L)^2/2$, then $x=y$.
\end{enumerate}
In particular, each coset of $2L$ in $L$ contains at most one closest vector of $L$ to $t$.
\end{lem}

\begin{lem}\label[lem]{lem:euclid-contraction}
Let $y$ be a closest vector of $L$ to $t$, let $\Delta>0$, and let $x\in y+2L$ satisfy $x\ne y$ and $\|x-t\|^2\le\|y-t\|^2+\Delta$. Let $\mathcal D$ be a distribution on $2L$ satisfying~\eqref{eq:dom} with $K=2L$, $H=I$, $r=2\sqrt\Delta$ and some $\gamma\in(0,1]$, and let $V\sim\mathcal D$. Then
\[
 \Pr\bigl[\|x+V-t\|^2\le\|y-t\|^2+\Delta/4\bigr]\ \ge\ \tfrac12\cdot21^{-N}\gamma.
\]
\end{lem}

For the proof, put $\eta=y-x$, $S=2L\cap\mathcal B(\sqrt{\Delta/6})$, and $E=\eta+S$. \Cref{lem:parity-geometry} makes every vector of $E$ improving and places $E$ inside the sampled ball, while \Cref{lem:cover,lem:count} give $|2L\cap\mathcal B(2\sqrt\Delta)|\le21^N|S|$. Thus at least a $\frac12 21^{-N}$ fraction of the relevant nonzero vectors are good; see \Cref{sec:euclid-deferred}.

\subsection{Algorithm for Exact Euclidean CVP}\label{sec:euclid-alg}
\Cref{alg:parity} presents our algorithm for exact Euclidean CVP. It processes the $2^N$ cosets of $2L$ separately and uses stages with excess bounds decreasing by a factor of four.
Put
\[
 U=(N+1)(\bar t^TG\bar t+\Tr G),\qquad \Delta_j=U/4^j,\qquad J=\min\{j\ge0:\Delta_j<2\iota(G)\}.
\]
Here, $U$ bounds the squared distance from $t$ to every starting point $Ba$, $a\in\{0,1\}^N$, and $\lambda_1(2L)^2\ge4\iota(G)$, so the final scale forces exactness.
Here, $J\ge1$ since $U\ge2\Tr G\ge2\iota(G)$.
In stage $j$ the sampler is called on $(4G,4\Delta_j,4\iota(G))$. With
\[
 \theta=\frac{189^{-N}}{1024N(\lceil\log_2(U/\iota(G))\rceil+4)^2},\qquad
 \mathcal M=\theta^{-1}\Bigl\lceil\log_2\frac{4\cdot2^NJ}{\varepsilon}\Bigr\rceil,
\]
we have $J=\poly(\ell)$ and $\mathcal M=2^{O(N)}\poly(\ell)$.

\begin{algorithm}[t]
\DontPrintSemicolon
\Input{Rational Gram matrix $G$, target coefficients $\bar t$, rational $\varepsilon\in(0,1/2)$}
\Output{Coefficient vector of a closest lattice point in the Euclidean norm}
\vspace{2mm}
$c_{\rm best}\gets0$\;
\For{$a\in\{0,1\}^N$}{
 $c\gets a$\;
 \For{$j=0,\ldots,J-1$}{
  $c_{\rm old}\gets c$\;
  \For{$\mathcal M$ times}{
   $\bar c\gets$ output of \Cref{alg:sampler} on $\bigl(4G,\,4\Delta_j,\,4\iota(G)\bigr)$\;
   \lIf{$\|c_{\rm old}+2\bar c-\bar t\|_G^2<\|c-\bar t\|_G^2$}{$c\gets c_{\rm old}+2\bar c$}
  }
 }
 \lIf{$\|c-\bar t\|_G^2<\|c_{\rm best}-\bar t\|_G^2$}{$c_{\rm best}\gets c$}
}
\Return{$c_{\rm best}$}\;
\caption{Exact CVP in the Euclidean norm}
\label{alg:parity}
\end{algorithm}

\paragraph{Correctness and resources.}\label{sec:euclid-analysis}
For the coset containing a closest vector $y$, maintain $\|x_j-t\|^2\le\|y-t\|^2+\Delta_j$. By \Cref{lem:euclid-contraction,thm:sampler}, each trial reaches the next scale with conditional probability at least $\theta$; repetition and a union bound over stages and cosets give the required success probability, and the last scale forces $x_J=y$. The algorithm makes $2^NJ\mathcal M=2^{O(N)}\poly(\ell)$ rank-$N$ oracle calls, and all stored quantities and queries have polynomial bit length, so the non-oracle space is polynomial. The two-run all-closest-vector variant and the bounded-error-oracle substitution use \Cref{lem:parity-geometry,lem:bounded-error}. Hence, \Cref{thm:main} holds for $p=2$; full details are in \Cref{sec:euclid-deferred}.

%% file: 5_lp_cvp.tex
\section{Exact CVP in \texorpdfstring{$\ell_p$}{lp} Norms}\label{sec:even}

We now prove \Cref{thm:main} for an integer $p\ge3$. The reduction keeps the coset-by-coset structure of \Cref{alg:parity}, but measures corrections in a diagonal metric computed from the current point. Throughout the section, $L=B\Z^N$ for a nonsingular $B\in\Q^{N\times N}$ and $t\in\Q^N$. Write
\[
 F_t(z)=\sum_{i=1}^N|z_i-t_i|^p,
 \qquad
 \phi(\alpha)=|\alpha|^p .
\]
A closest vector in the $\ell_p$ norm is a minimizer of $F_t$ on $L$.

\subsection{Midpoint Inequalities}\label{sec:midpoint}

Let $y$ be a minimizer and let $x\in y+2L$. Since $(x+y)/2\in L$, minimality of $y$ gives a midpoint inequality. The following bounds are the quantitative form needed below; the proof is deferred to \Cref{sec:lp-deferred}.

\begin{lem}\label[lem]{lem:even-midpoint}
Let $y$ be a minimizer of $F_t$ on $L$, let $\Delta>0$, and let $x\in y+2L$ satisfy $F_t(x)\le F_t(y)+\Delta$. Write $\eta=y-x$, $\sigma=x-t$ and $\sigma'=y-t$. Then,
\[
 \sum_i |\eta_i|^p\le2^p\Delta,
 \qquad \sum_i \max\{|\sigma_i|,|\sigma'_i|\}^{p-2}\eta_i^2\le4\Delta,
 \qquad \sum_i\bigl(|\sigma'_i|^{p/2}-|\sigma_i|^{p/2}\bigr)^2\le2\Delta.
\]
\end{lem}

\subsection{A Weighted Norm Computed at the Current Point}\label{sec:weighted}

The local curvature of $F_t$ at the unknown minimizer $y$ is unavailable to the algorithm. We instead use weights computed from the current point $x$. For $\Delta>0$, put $\beta=(\Delta/N)^{1/p}$ and let $\rho$ be the least number of the form $2^k$, $k\in\Z$, such that $\rho^p\ge(\Delta/N)^{p-2}$. Thus $\beta^{p-2}\le\rho<2\beta^{p-2}$. Define
\begin{equation}\label{eq:even-regularizer}
 h_i=|x_i-t_i|^{p-2}+\rho,
 \qquad H_x=\operatorname{diag}(h_1,\ldots,h_N),
 \qquad \beta_i=\max\{|y_i-t_i|,\beta\}.
\end{equation}
The metric $H_x$ is rational and computable from $x$ and $\Delta$; the numbers $\beta_i$ are used only in the analysis.

\begin{lem}\label[lem]{lem:weighted}
Let $y$ be a minimizer of $F_t$ on $L$, let $\Delta>0$, and let $x\in y+2L$ satisfy $F_t(x)\le F_t(y)+\Delta$. With~\eqref{eq:even-regularizer},
\[
 \|y-x\|_{H_x}^2\le12\Delta,
 \qquad
 \sum_i\frac{\beta_i^{p-2}}{h_i}\le7N,
 \qquad
 h_i\beta_i^2\ge\frac\Delta N\quad\text{for every }i.
\]
\end{lem}

The first bound places the unknown correction $y-x$ inside a known $H_x$-ball. The other two bounds compare the computable weights at $x$ with the curvature around $y$. The proof is given in \Cref{sec:lp-deferred}.

\subsection{Progress in the Weighted Norm}\label{sec:lp-progress}

We use the following coordinatewise Taylor bound; its proof is also deferred to  \Cref{sec:lp-deferred}.

\begin{lem}\label[lem]{lem:taylor}
For $a,w\in\R$ and $\beta_0>0$ with $\beta_0\ge|a|$,
\[
 0\ \le\ \phi(a+w)-\phi(a)-\phi'(a)w
 \ \le\ \tfrac{p^2}{2}\beta_0^{p-2}e^{p|w|/\beta_0}w^2 .
\]
\end{lem}

The next lemma is the analogue of \Cref{lem:euclid-contraction}. It is the only new geometric ingredient needed to reuse the Euclidean iteration.

\begin{lem}\label[lem]{lem:lp-contraction}
Let $y$ be a minimizer of $F_t$ on $L$, let $\Delta>0$, and let $x\in y+2L$ satisfy $x\ne y$ and $F_t(x)\le F_t(y)+\Delta$. Let $H_x$ be defined by~\eqref{eq:even-regularizer}, let $\mathcal D$ be a distribution on $2L$ satisfying~\eqref{eq:dom} with $K=2L$, $H=H_x$, $r=4\sqrt\Delta$ and some $\gamma\in(0,1]$, and let $V\sim\mathcal D$. Then
\[
 \Pr\bigl[F_t(x+V)\le F_t(y)+\Delta/4\bigr]
 \ \ge\ \tfrac14\cdot(188p)^{-N}\gamma.
\]
\end{lem}

\begin{proof}[Proof sketch]
Let $\eta=y-x$, $K=2L$, and $Q=\{w:h_iw_i^2\le\Delta/(32Np^2)~\text{for all}~i\}$. Keep the half of $K\cap Q$ on which the linear term $\sum_i\phi'(y_i-t_i)w_i$ is nonpositive. Symmetry preserves at least half the points, and \Cref{lem:weighted,lem:taylor} show that every retained $w$ gives $F_t(y+w)-F_t(y)<\Delta/4$ and that $\eta+w$ lies in $\mathcal B_{H_x}(4\sqrt\Delta)$. After the map $w\mapsto H_x^{1/2}w$, \Cref{lem:count,lem:cover} show that this box contains, up to the factor $(188p)^N$, enough points of the sampled ball. Hence at least a $\frac14(188p)^{-N}$ fraction of its relevant nonzero vectors are improving, and~\eqref{eq:dom} gives the lemma. See \Cref{sec:lp-deferred} for the coordinate estimates and covering calculation.
\end{proof}

\paragraph{Reduction and completion of the proof.}\label{sec:lp-alg}
Use \Cref{alg:parity}, but at the beginning of stage $j$ freeze $H=H_{x_{\rm old}}$ from~\eqref{eq:even-regularizer} and use this metric for every sample in that stage. If $d$ is the least common multiple of the denominators of $B$ and $t$, put
\[\textstyle
 U=\sum_i\Bigl(|t_i|+\sum_k|B_{ik}|\Bigr)^p,\qquad
 \Delta_j=U/4^j,\qquad J=\min\{j\ge0:\Delta_j<d^{-p}\}.
\]
Then $F_t(Ba)\le U$ for every starting coset representative, while $F_t(L)\subset d^{-p}\Z$, so the last scale forces exact optimality. 
Also $J\ge1$, since nonsingularity of $B$ gives $U\ge Nd^{-p}$.
In stage $j$ the sampler input and the uniform success bound are
\[
 \bigl((2B)^TH(2B),16\Delta_j,4\rho/d^2\bigr),\qquad
 \theta=\frac{(1692p)^{-N}}{2048N(\lceil\log_2(4Ud^{p+2}N)\rceil+4)^2},
\]
and we take $\mathcal M=\theta^{-1}\lceil\log_2(4\cdot2^NJ/\varepsilon)\rceil=(1692p)^N\poly(p,\ell)$. By \Cref{lem:lp-contraction,thm:sampler}, repetition preserves the invariant $F_t(x_j)\le F_t(y)+\Delta_j$ 
through all stages of each coset containing a minimizer
with the required probability. All metrics, coefficient vectors, and queries have $\poly(p,\ell)$ bits, giving $(Cp)^N\poly(p,\ell)$ time outside the oracle and polynomial working space. Hence \Cref{thm:main} holds for every integer $p\ge3$; full pseudocode, resource bounds, the all-closest-vector variant, and the bounded-error-oracle argument are in \Cref{sec:lp-deferred}.

%% file: 6_sampler_proof.tex
\section{Proof of the Short-Vector Sampler Theorem}
\label{sec:sampler}

In this section, we prove \Cref{thm:sampler} for \Cref{alg:sampler}. Let $\mathcal D$ denote the distribution of the lattice vector represented by the algorithm's output coefficient vector, and let $V\sim\mathcal D$. Fix a nonzero vector $x=mu$, where $m\ge1$ and $u$ is primitive. The proof has three steps. We first establish counting bounds for lattice points in balls. We then show that a suitable random sparsification isolates $\pm u$ with sufficient probability. Finally, we combine these bounds with the random choices of \Cref{alg:sampler} to prove the pointwise probability bound~\eqref{eq:dom} and the claimed resource bounds.

Throughout this section, $K=B_K\Z^N$, $H$ and $G_K=B_K^THB_K$ are as in \Cref{sec:sampler-statement}. All norms and balls are taken with respect to $H$, and we drop $H$ from the notation. We write $\lambda=\lambda_1(K,H)$ and $\xi=\xi_{K,H}$. For a lattice vector $u$, we write $a_u$ for its coefficient vector.

\paragraph{Basis for the Sparsified Sublattices.}
For a prime $q$ and $z\in\F_q^N$, recall
\[
 \Lambda_z=\{a\in\Z^N:\ip za\equiv0\pmod q\},\qquad K_z=B_K\Lambda_z .
\]
If $z=0$, take $D_z=I_N$. If $z\ne0$, fix a coordinate $l_0$ with $z_{l_0}\ne0$ and let $\bar z_l\in\{0,\ldots,q-1\}$ represent $z_lz_{l_0}^{-1}$ for $l\ne l_0$. Taking $qe_{l_0}$ and $e_l-\bar z_le_{l_0}$, $l\ne l_0$, as columns gives an integer matrix $D_z$ with $|\det D_z|=q$ whose columns form a basis of $\Lambda_z$. Hence, $K_z$ has Gram matrix $D_z^TG_KD_z$, whose bit length is polynomial in that of $G_K$ and in $\log q$, and an oracle answer $a'$ for $K_z$ corresponds to the coefficient vector $D_za'$ in $K$.

\paragraph{Relation to Stephens--Davidowitz.}
The sparsification step follows the primitive-vector sampler of Stephens-Davidowitz~\cite[Section~4.1]{Ste16full}, which builds on Khot~\cite{Kho05} and Dadush and Kun~\cite{DK16}. The results of Stephens-Davidowitz~\cite[Lemma~2.18, Theorem~4.1, Proposition~4.2]{Ste16full} isolate a primitive direction under conditions involving the number of short primitive vectors and a lower bound on $\lambda_1(K,H)$. The lemmas below replace these conditions by the single inequality $q>|K\cap\mathcal B_H(2s)|$ for $s\ge\|u\|_H$, counting all lattice points and absorbing the loss through \Cref{lem:multiples}. We also guess the order of magnitude of $q$ instead of estimating the number of short primitive vectors. Finally, unlike the discrete-Gaussian sampler of~\cite[Theorem~4.6]{Ste16full}, our construction needs only the centered pointwise lower bound~\eqref{eq:dom}; no shifted sampler is used.

\subsection{Counting Lattice Points in Balls}\label{sec:counting}

The first lemma bounds the number of lattice points in a ball in terms of the minimum distance, and compares balls whose radii differ by a factor of two.

\begin{lem}\label[lem]{lem:packing}
For every $R>0$, $|K\cap \mathcal B(R)|\le(1+2R/\lambda)^N$. For every $s>0$, $|K\cap \mathcal B(2s)|\le9^N|K\cap \mathcal B(s)|$.
\end{lem}
\begin{proof}
Distinct points of $K$ are at distance at least $\lambda$. Hence the open balls of radius $\lambda/2$ centered at the points of $K\cap \mathcal B(R)$ are pairwise disjoint, and they are contained in $\mathcal B(R+\lambda/2)$. Comparing volumes gives $|K\cap \mathcal B(R)|\,(\lambda/2)^N\le(R+\lambda/2)^N$, which is the first bound. For the second bound, \Cref{lem:cover}(i) with $R=2s$ and $\delta=s/2$ covers $\mathcal B(2s)$ by $9^N$ translates of $\mathcal B(s/2)$, and \Cref{lem:count} with $Q=\mathcal B(s)$ gives the claim.
\end{proof}

The second lemma compares the number of lattice points in a ball of radius $r$ with the number in a ball of radius $r/2^\kappa$. Every primitive vector $u$ of length at most $r/2^\kappa$ has the signed multiples $\pm u,\ldots,\pm2^\kappa u$ in the larger ball, and \Cref{lem:multiples} makes this count precise when $u$ ranges over all primitive vectors.

\begin{lem}\label[lem]{lem:multiples}
Let $r>0$, let $\kappa\ge0$ be an integer, and let $s=r/2^\kappa$. Then,
\[
 |K\cap \mathcal B(r)|-1\ \ge\ 2^\kappa\bigl(|K\cap \mathcal B(s)|-1\bigr).
\]
\end{lem}
\begin{proof}
Since every nonzero vector is $mu$ for a unique $m\ge1$ and a unique primitive $u$,
\[
 |K\cap \mathcal B(R)|-1\ =\ 2\sum_{m\ge1}\xi(R/m)\qquad\text{for every }R>0 .
\]
For $m\ge1$, let $\mathcal I_m=\{k\in\Z:2^\kappa(m-1)<k\le2^\kappa m\}$. The sets $\mathcal I_m$ partition the positive integers into blocks of $2^\kappa$ consecutive integers. Since $\xi$ is nondecreasing, every $k\in \mathcal I_m$ satisfies $\xi(r/k)\ge\xi(r/(2^\kappa m))=\xi(s/m)$. Therefore,
\[
 \sum_{k\ge1}\xi(r/k)\ =\ \sum_{m\ge1}\sum_{k\in \mathcal I_m}\xi(r/k)\ \ge\ 2^\kappa\sum_{m\ge1}\xi(s/m),
\]
and the lemma follows from the identity above.
\end{proof}

\subsection{Sparsification}\label{sec:sparsification}

We use the sublattices $K_z$ defined in \Cref{sec:sampler-statement}. To isolate a primitive vector $u$, the other short primitive vectors must be independent of $u$ modulo $q$, and no multiple of a removed primitive vector may be shorter than $u$. \Cref{lem:nonprop,lem:isolation} establish both properties under the single condition $q>|K\cap \mathcal B(2s)|$ for a radius $s\ge\|u\|$. They are the variant of~\cite[Proposition~4.2]{Ste16full} described in \Cref{rem:ste16}.

\begin{lem}\label[lem]{lem:nonprop}
Let $s>0$ and let $q$ be a prime with $q>|K\cap \mathcal B(2s)|$. If $u_1,u_2\in K\cap \mathcal B(s)$ are primitive and $u_2\ne\pm u_1$, then the coefficient vectors of $u_1$ and $u_2$ are linearly independent over $\F_q$.
\end{lem}
\begin{proof}
We show that a dependence modulo $q$ forces at least $q$ lattice points into the parallelogram spanned by $u_1$ and $u_2$, which lies in $\mathcal B(2s)$.
The coefficient vectors $a_{u_1}$ and $a_{u_2}$ have greatest common divisor $1$, so neither vanishes modulo~$q$. Suppose that they are linearly dependent over $\F_q$. Then $a_{u_2}\equiv k_0a_{u_1}\pmod q$ for some integer $k_0$, that is, $u_2-k_0u_1\in qK$. Two primitive vectors on a common line through the origin are equal up to sign, so $u_1$ and $u_2$ are linearly independent over $\R$.
Thus, $K_{u_1,u_2}=K\cap\spn(u_1,u_2)$ is a lattice of rank two, in which $u_1$ is primitive. Since $qK\cap\spn(u_1,u_2)=qK_{u_1,u_2}$, the vector $u_2$ lies in the sublattice $K'=\Z u_1+qK_{u_1,u_2}$.

We compute the index of $K'$. The quotient $K_{u_1,u_2}/qK_{u_1,u_2}\cong\F_q^2$ has $q^2$ elements, and the image of $K'$ in it is the subgroup generated by the image of $u_1$. This image is nonzero because $u_1$ is primitive in $K_{u_1,u_2}$, so the subgroup has $q$ elements and $[K_{u_1,u_2}:K']=q$. Since $\Z u_1+\Z u_2\subseteq K'$, it follows that $[K_{u_1,u_2}:\Z u_1+\Z u_2]\ge q$.

The half-open parallelogram $\{\alpha_1u_1+\alpha_2u_2:\alpha_1,\alpha_2\in[0,1)\}$ is a fundamental domain of $\Z u_1+\Z u_2$, so it contains exactly $[K_{u_1,u_2}:\Z u_1+\Z u_2]\ge q$ points of $K_{u_1,u_2}$. Each of them has norm less than $\|u_1\|+\|u_2\|\le2s$. Hence $|K\cap \mathcal B(2s)|\ge q$, contradicting the choice of $q$.
\end{proof}

\begin{lem}\label[lem]{lem:isolation}
Let $s>0$, let $q$ be a prime with $q>|K\cap \mathcal B(2s)|$, and let $u\in K$ be primitive with $\|u\|\le s$. If $z$ is uniform in $\F_q^N$, then with probability at least $1/(2q)$ the shortest nonzero vectors of $K_z$ are exactly $u$ and $-u$.
\end{lem}
\begin{proof}
We first bound the probability that $u$ is, up to sign, the only primitive vector of $\mathcal B(s)$ in $K_z$. We then show that in this case no other nonzero vector of $K_z$ is as short as $u$.

The vector $0$ and the two vectors in each of the $\xi(s)$ primitive pairs in $\mathcal B(s)$ are distinct points of $K\cap\mathcal B(2s)$.
Hence, $2\xi(s)<q$. 
Let $\mathcal E$ be the event that $u\in K_z$ and that $u'\notin K_z$ for every primitive $u'\in\mathcal B(s)$ with $u'\ne\pm u$.
Choose one representative $u'$ of each of the $\xi(s)-1$ other primitive pairs in $\mathcal B(s)$. By \Cref{lem:nonprop}, the coefficient vectors of $u$ and $u'$ are linearly independent over $\F_q$, so a union bound gives
\[
 \Pr[\mathcal E]\ \ge\ \Pr[u\in K_z]-\sum_{u'}\Pr[u\in K_z\text{ and }u'\in K_z]\ =\ \frac1q-\frac{\xi(s)-1}{q^2}\ \ge\ \frac1q-\frac1{2q}\ =\ \frac1{2q}.
\]

Now assume $\mathcal E$, and let $w\in K_z$ be nonzero with $\|w\|\le\|u\|$. Write $w=kw'$ with $k\ge1$ and $w'$ primitive. Since $\|w'\|\le\|w\|\le\|u\|\le s$, we have $w'\in\mathcal B(s)$. If $w'\in K_z$, then $w'=\pm u$ by the definition of $\mathcal E$, and $\|w\|\le\|u\|$ forces $w=\pm u$. If $w'\notin K_z$, then $\ip z{a_{w'}}\not\equiv0$ while $k\ip z{a_{w'}}\equiv0\pmod q$, so $q$ divides $k$ and $\|w\|\ge q\lambda$. The multiples of a shortest nonzero vector show that $|K\cap \mathcal B(2s)|\ge1+2\lfloor2s/\lambda\rfloor>2s/\lambda$, so $q\lambda>2s>\|u\|$, which is impossible. Hence on $\mathcal E$ the shortest nonzero vectors of $K_z$ are exactly $u$ and $-u$.
\end{proof}

\subsection{Analysis of \Cref{thm:sampler}}\label{sec:sampler-proof}

We begin by restating \Cref{thm:sampler} for convenience.

\begin{restatedstatement}{thm:sampler}
Let $K=B_K\Z^N$ and let $H\succ0$ be a metric such that $G_K:=B_K^THB_K\in\Q^{N\times N}$.
Let $r^2>0$ and $\lambda_0^2>0$ be rational numbers satisfying $\lambda_0\le\lambda_1(K,H)$. There is a randomized algorithm with access to an exact SVP oracle that, given $G_K$, $r^2$, and $\lambda_0^2$, outputs a coefficient vector representing a lattice vector in $K$. Let $\mathcal D$ be the induced distribution of this lattice vector, and let $V\sim\mathcal D$. Then the following properties hold.
\begin{enumerate}[label=(\alph*)]
\item\label{it:support} $\operatorname{supp}(\mathcal D)\subseteq K\cap \mathcal B_H(r)$; equivalently, $V\in K\cap \mathcal B_H(r)$ with probability $1$.
\item\label{it:dom} $\mathcal D$ satisfies~\eqref{eq:dom} with
\[
 \gamma=\frac{9^{-N}}{512\,N\,\bigl(\log_2\max\{1,r/\lambda_0\}+4\bigr)^2},
\]
conditionally on any history of earlier computations and for every rule by which the oracle chooses among shortest vectors.
\item\label{it:res} The algorithm makes at most one oracle call, on a full-rank sublattice of $K$ in the metric $H$ given by a Gram matrix whose bit length is polynomial in that of the input. Its other bit operations and its space are polynomial in the bit lengths of the input and of the oracle answer.
\end{enumerate}
\end{restatedstatement}

We first describe the random sampling procedures used by
\Cref{alg:sampler}, and then prove the three guarantees of
\Cref{thm:sampler}.

\paragraph{Random choices.}
Lines~\ref{ln:guess} and~\ref{ln:sparse} of \Cref{alg:sampler} use the following procedures.
\begin{enumerate}[label=(\arabic*)]
\item To sample from a set $\{a_0,\ldots,a_0+n-1\}$ of $n$ consecutive integers using a bounded number of random bits, the procedure draws $\lceil\log_2n\rceil$ random bits and reads them as an integer $\nu$. It returns $a_0+\nu$ if $\nu<n$, and returns \textsc{fail} otherwise. Conditioned on success, the returned element is uniform over the set. Unconditionally, each element is returned with probability $2^{-\lceil\log_2n\rceil}\ge1/(2n)$, which equals $1/n$ when $n$ is a power of $2$.
\item A prime in $[2^i,2^{i+1})$, $i\ge1$, is obtained by drawing at most $3(i+1)$ uniform integers from this interval and returning the first prime among them, where primality is tested deterministically~\cite{AKS04}. By~\cite[Theorem~5.11]{Shoup03beta}, $\pi(2m)-\pi(m)>m/(3\log(2m))$ for $m\ge2$. With $m=2^i$, a uniform integer from $[2^i,2^{i+1})$ is prime with probability greater than $1/(3(i+1))$, so a prime is returned with probability greater than $1-e^{-1}>1/2$.
\item A uniform vector $z\in\F_q^N$, for a prime $q\in[2^i,2^{i+1})$, is obtained coordinate by coordinate. For each coordinate the procedure draws up to $\lceil\log_2N\rceil+2$ integers of $i+1$ bits and keeps the first that is less than $q$. Each draw is kept with probability at least $1/2$, so a given coordinate fails with probability at most $1/(4N)$, and all coordinates are obtained with probability at least $3/4$. Conditioned on this event, $z$ is uniform on $\F_q^N$.
\end{enumerate}

\paragraph{Support (proof of Theorem~\ref{thm:sampler}\ref{it:support}).}
In any execution, the output is $0$ or a coefficient vector $\pm ma$ representing the realized lattice vector $v=B_K(\pm ma)$. The output test ensures
\[
 \|v\|_H^2=m^2a^TG_Ka\le r^2,
\]
so $v\in K\cap\mathcal B_H(r)$. Thus, $\operatorname{supp}(\mathcal D)\subseteq K\cap\mathcal B_H(r)$, proving part~\ref{it:support} of Theorem~\ref{thm:sampler}.

\paragraph{Pointwise lower bound (proof of Theorem~\ref{thm:sampler}\ref{it:dom}).}
Fix a nonzero $x\in K\cap \mathcal B(r)$, so that $r\ge\lambda\ge\lambda_0$. We identify random choices under which \Cref{alg:sampler} returns $x$, bound the product of their probabilities, and compare it with $1/|K\cap \mathcal B(r)|$.

Write $x=mu$ with $m\ge1$ and $u$ primitive, and let $\kappa=\lfloor\log_2m\rfloor$, $s=r/2^\kappa$ and $\Xi=|K\cap \mathcal B(2s)|$. Since $2^\kappa\le m\le r/\|u\|\le r/\lambda_0$, we have $4^\kappa\lambda_0^2\le r^2$, that is, $\kappa\le J_{\rm s}$, as defined in Line~\ref{ln:js} of \Cref{alg:sampler}. Since $m\ge2^\kappa$, we have $\|u\|\le r/m\le s$, and in particular $s\ge\lambda$. The next claim shows that some exponent $i$ in the range of the algorithm yields primes for which \Cref{lem:isolation} applies.

\begin{claim}\label[claim]{cl:scale}
Let $i^*$ be the least integer $i\ge1$ with $2^i>\Xi$. Then $i^*\le i_{\max}$, and every prime $q\in[2^{i^*},2^{i^*+1})$ satisfies $\Xi<q\le4\Xi$.
\end{claim}
\begin{proof}
By \Cref{lem:packing},
$\Xi\le(1+4s/\lambda)^N
\le(1+2^{J_{\rm s}+3})^N
\le2^{N(J_{\rm s}+4)}
<2^{i_{\max}}$,
where we used $s\le r<2^{J_{\rm s}+1}\lambda_0\le2^{J_{\rm s}+1}\lambda$. Hence, $i^*\le i_{\max}$. Since $K\cap\mathcal B(2s)$ contains $0$ and $\pm u$, we have $\Xi\ge3$, so $i^*\ge2$ and $2^{i^*-1}\le\Xi$. Therefore, $\Xi<2^{i^*}\le q<2^{i^*+1}\le4\Xi$.
\end{proof}

The algorithm returns $x$ if the following events occur. Each probability bound holds conditionally on the preceding events and on any earlier history.
\begin{enumerate}[label=(E\arabic*),nosep]
\item Procedure (1) returns this $\kappa$, $i=i^*$ and this $m$: probability at least $\frac1{2(J_{\rm s}+1)}\cdot\frac1{2i_{\max}}\cdot2^{-\kappa}$, since the range of $m$ has $2^\kappa$ elements.
\item Procedures (2) and (3) succeed: probability at least $\frac12\cdot\frac34$. Then $z$ is uniform on $\F_q^N$.
\item The shortest nonzero vectors of $K_z$ are exactly $\pm u$: probability at least $\frac1{2q}\ge\frac1{8\Xi}$, by \Cref{lem:isolation} and \Cref{cl:scale}. On this event the oracle returns a coefficient vector $a'$ with $D_za'=\pm a_u$ for every rule of choosing among shortest vectors, and the output test passes because $\|mu\|\le r$.
\item The random sign of the output is the one that gives $a_x$: probability $\frac12$.
\end{enumerate}
Multiplying these bounds gives
\[
 \mathcal D(x)\ \ge\ \frac{3\cdot2^{-\kappa}}{512\,(J_{\rm s}+1)\,i_{\max}\,\Xi}.
\]
It remains to compare $2^\kappa\Xi$ with $|K\cap \mathcal B(r)|$. By \Cref{lem:packing}, the inequality $|K\cap \mathcal B(s)|\ge3$, and \Cref{lem:multiples},
\[
 2^\kappa\Xi\ \le\ 9^N2^\kappa|K\cap \mathcal B(s)|\ \le\ \tfrac32\,9^N2^\kappa\bigl(|K\cap \mathcal B(s)|-1\bigr)\ \le\ \tfrac32\,9^N|K\cap \mathcal B(r)|.
\]
Finally, $(J_{\rm s}+1)i_{\max}\le2N(J_{\rm s}+4)^2$ and $J_{\rm s}\le\log_2(r/\lambda_0)=\log_2\max\{1,r/\lambda_0\}$. Combining the three displays,
\[
 \mathcal D(x)\ \ge\ \frac{9^{-N}}{256\,(J_{\rm s}+1)\,i_{\max}\,|K\cap \mathcal B(r)|}\ \ge\ \frac{9^{-N}}{512\,N\,\bigl(\log_2\max\{1,r/\lambda_0\}+4\bigr)^2\,|K\cap \mathcal B(r)|},
\]
which is~\eqref{eq:dom} with the value of $\gamma$ in \Cref{thm:sampler}\ref{it:dom}. This proves part~\ref{it:dom}.

\paragraph{Resources (proof of Theorem~\ref{thm:sampler}\ref{it:res}).}
If $r^2<\lambda_0^2$, the algorithm returns $0$ before making an oracle call, and the resource bounds are immediate. Hence assume $r^2\ge\lambda_0^2$. The algorithm makes at most one oracle call, on the full-rank sublattice $K_z$ of $K$, which is $K$ itself for $z=0$. Since $4^{J_{\rm s}}\le r^2/\lambda_0^2$, the numbers $J_{\rm s}$ and $i_{\max}$ are polynomial in the input length, and so are the bit lengths of $m$, $q$, $z$ and $D_z$, the number of random bits, and the number of primality tests. 
In particular, $q<2^{i_{\max}+1}$ has at most $i_{\max}+1=N(J_{\rm s}+4)+2$ bits, by the definition of $i_{\max}$. Since the entries of $D_z$ have $O(i_{\max})$ bits and $G_K$ is a rational input matrix, standard rational matrix multiplication shows that every entry of the query $D_z^TG_KD_z$ has bit length polynomial in the input length.
All remaining computations involve rationals whose bit length is polynomial in the bit lengths of the input and of the oracle answer. This proves part~\ref{it:res} and completes the proof of \Cref{thm:sampler}.

%% file: 7_deferred_cvp_proofs.tex
\section{Deferred Proofs and Details for the CVP Reductions}\label{sec:cvp-deferred}

This section contains the proofs and detailed bookkeeping deferred from \Cref{sec:setup,sec:euclidean,sec:even}. After the Euclidean proofs, we record polynomial-space consequences of the CVP reductions before turning to the deferred $\ell_p$ proofs.

\subsection{Counting and Covering Lemmas}

We prove \Cref{lem:count,lem:cover}, restating them for convenience.

\begin{restatedstatement}{lem:count}
Let $L\subset\R^N$ be a lattice and $Q\subset\R^N$ a convex, compact, symmetric set. If $X\subseteq\bigcup_{k=1}^T(c_k+\frac12Q)$ for some $c_1,\ldots,c_T\in\R^N$, then $|L\cap X|\le T\cdot|L\cap Q|$.
\end{restatedstatement}

\begin{proof}[Proof of \Cref{lem:count}]
Fix $k$ such that $L\cap(c_k+\frac12Q)$ contains a point $x_k$. For every $x$ in this set, $x-x_k\in\frac12Q-\frac12Q$, and $\frac12Q-\frac12Q=Q$ because $Q$ is convex and symmetric. As $x-x_k\in L$, we obtain $L\cap(c_k+\frac12Q)\subseteq x_k+(L\cap Q)$. Summing over $k$ proves the lemma.
\end{proof}

\begin{restatedstatement}{lem:cover}
The following covering bounds hold.
\begin{enumerate}[label=(\roman*)]
\item For every positive definite metric $H$ and $R,\delta>0$, the ball $\mathcal B_H(R)$ is contained in a union of at most $(1+2R/\delta)^N$ translates of $\mathcal B_H(\delta)$.
\item For the Euclidean norm, $R>0$ and $0<\alpha\le R/\sqrt N$, the ball $\mathcal B(R)$ is contained in a union of at most $\omega_N\cdot(2R/\alpha)^N$ translates of the cube $[0,\alpha]^N$.
\end{enumerate}
\end{restatedstatement}

\begin{proof}[Proof of \Cref{lem:cover}]
(i) Let $Y\subseteq\mathcal B_H(R)$ be maximal subject to $\|x-x'\|_H>\delta$ for distinct $x,x'\in Y$. The balls $x+\mathcal B_H(\delta/2)$, $x\in Y$, are disjoint and contained in $\mathcal B_H(R+\delta/2)$, so $|Y|\le(1+2R/\delta)^N$. By maximality, $\mathcal B_H(R)\subseteq\bigcup_{x\in Y}(x+\mathcal B_H(\delta))$.

(ii) The cubes $\alpha(k+[0,1)^N)$, $k\in\Z^N$, are disjoint and have diameter $\alpha\sqrt N\le R$. Each cube that meets $\mathcal B(R)$ is contained in $\mathcal B(2R)$, so at most $\omega_N\cdot(2R/\alpha)^N$ cubes meet $\mathcal B(R)$, and their closures cover $\mathcal B(R)$.
\end{proof}

\subsection{Deferred Euclidean Proofs}\label{sec:euclid-deferred}
We first prove \Cref{lem:parity-geometry,lem:euclid-contraction}. We then give the detailed correctness and resource analysis of \Cref{alg:parity} and complete the proof of \Cref{thm:main} for $p=2$.

\begin{restatedstatement}{lem:parity-geometry}
Let $y$ be a closest vector of $L$ to $t$. For every $w\in2L$,
\[
 \tfrac12\|w\|^2\ \le\ \|y+w-t\|^2-\|y-t\|^2\ \le\ \tfrac32\|w\|^2 .
\]
Consequently, for every $\Delta\ge0$ and every $x\in y+2L$ with $\|x-t\|^2\le\|y-t\|^2+\Delta$,
\begin{enumerate}[label=(\roman*),nosep]
\item\label{it:geo-close} $\|x-y\|^2\le2\Delta$;
\item\label{it:geo-equal} if $\Delta<\lambda_1(2L)^2/2$, then $x=y$.
\end{enumerate}
In particular, each coset of $2L$ in $L$ contains at most one closest vector of $L$ to $t$.
\end{restatedstatement}

\begin{proof}[Proof of \Cref{lem:parity-geometry}]
Since $w\in2L$, both $y+w/2$ and $y-w/2$ lie in $L$. As $y$ is a closest vector, neither of them is closer to $t$ than $y$, and expanding the squares gives
\[
 0\ \le\ \|y\pm w/2-t\|^2-\|y-t\|^2\ =\ \pm\ip{y-t}{w}+\tfrac14\|w\|^2 .
\]
Hence, $|\ip{y-t}{w}|\le\|w\|^2/4$. Inserting this bound into the identity $\|y+w-t\|^2-\|y-t\|^2=\|w\|^2+2\ip{y-t}{w}$ gives both inequalities.

For~\ref{it:geo-close}, apply the lower bound to $w=x-y\in2L$. For~\ref{it:geo-equal}, suppose that $x\ne y$. Then $x-y$ is a nonzero vector of $2L$, so $\|x-y\|^2\ge\lambda_1(2L)^2>2\Delta$, which contradicts~\ref{it:geo-close}. The last statement is~\ref{it:geo-equal} with $\Delta=0$.
\end{proof}

\begin{restatedstatement}{lem:euclid-contraction}
Let $y$ be a closest vector of $L$ to $t$, let $\Delta>0$, and let $x\in y+2L$ satisfy $x\ne y$ and $\|x-t\|^2\le\|y-t\|^2+\Delta$. Let $\mathcal D$ be a distribution on $2L$ satisfying~\eqref{eq:dom} with $K=2L$, $H=I$, $r=2\sqrt\Delta$ and some $\gamma\in(0,1]$, and let $V\sim\mathcal D$. Then
\[
 \Pr\bigl[\|x+V-t\|^2\le\|y-t\|^2+\Delta/4\bigr]\ \ge\ \tfrac12\cdot21^{-N}\gamma.
\]
\end{restatedstatement}

\begin{proof}[Proof of \Cref{lem:euclid-contraction}]
Let $\eta=y-x$, let $S=2L\cap \mathcal B(\sqrt{\Delta/6})$, and let $E=\eta+S$. We show that
\begin{enumerate}[label=(\alph*),nosep]
\item\label{it:E-good} $\|x+z-t\|^2\le\|y-t\|^2+\Delta/4$ for every $z\in E$,
\item\label{it:E-ball} $E\subseteq2L\cap \mathcal B(2\sqrt\Delta)$, and
\item\label{it:E-large} $|E\setminus\{0\}|\ge\frac12\cdot21^{-N}\,|2L\cap \mathcal B(2\sqrt\Delta)|$.
\end{enumerate}
Then~\eqref{eq:dom} gives $\Pr[V\in E]=\mathcal D(E)\ge\gamma\,|E\setminus\{0\}|/|2L\cap \mathcal B(2\sqrt\Delta)|\ge\frac12\cdot21^{-N}\gamma$, and~\ref{it:E-good} proves the lemma.

\begin{itemize}[nosep]
\item For~\ref{it:E-good}, let $z=\eta+w$ with $w\in S$. Then $x+z=y+w$, and the upper bound of \Cref{lem:parity-geometry} gives $\|y+w-t\|^2-\|y-t\|^2\le\frac32\|w\|^2\le\Delta/4$.

\item For~\ref{it:E-ball}, $\eta\in2L$ because $x\in y+2L$, and $\|\eta\|\le\sqrt{2\Delta}$ by \Cref{lem:parity-geometry}\ref{it:geo-close}. Hence, every $w\in S$ satisfies $\|\eta+w\|\le(\sqrt2+1/\sqrt6)\sqrt\Delta<2\sqrt\Delta$.

\item For~\ref{it:E-large}, \Cref{lem:cover}(i) with $R=2\sqrt\Delta$ and $\delta=\frac12\sqrt{\Delta/6}$ covers $\mathcal B(2\sqrt\Delta)$ by at most $(1+8\sqrt6)^N\le21^N$ translates of $\frac12\mathcal B(\sqrt{\Delta/6})$. By \Cref{lem:count} with $Q=\mathcal B(\sqrt{\Delta/6})$,
\[
 |2L\cap \mathcal B(2\sqrt\Delta)|\ \le\ 21^N|S| .
\]
\end{itemize}
Finally $|E|=|S|$, and $E$ contains $\eta\ne0$ because $0\in S$. Hence $|E\setminus\{0\}|\ge\max\{1,|S|-1\}\ge|S|/2$.
\end{proof}

\paragraph{Detailed probability and resource analysis.}
We provide the detailed correctness and resource analysis underlying the proof sketch in \Cref{sec:euclid-analysis}.
\begin{lem}\label[lem]{lem:euclid-coset}
Assume that the oracle is exact. Let $y$ be a closest vector of $L$ to $t$, and let $a\in\{0,1\}^N$ satisfy $y\in Ba+2L$. Then, with probability at least $1-\varepsilon/(4\cdot2^N)$, the iteration of \Cref{alg:parity} for $a$ ends with $Bc=y$.
\end{lem}
\begin{proof}
For $0\le j\le J$, let $x_j$ be the current point $Bc$ at the start of stage $j$, where $x_J$ denotes the point at the end of the last stage. All these points lie in $y+2L$, since the algorithm adds only vectors of $2L$ to $Ba$. We have $\|x_0-t\|^2\le U=\Delta_0\le\|y-t\|^2+\Delta_0$, so the invariant holds initially.

Fix $j<J$ and condition on the computation up to the start of stage $j$, assuming that $\|x_j-t\|^2\le\|y-t\|^2+\Delta_j$. We show that $\|x_{j+1}-t\|^2\le\|y-t\|^2+\Delta_{j+1}$ with probability at least $1-\varepsilon/(4\cdot2^NJ)$. If $x_j=y$, then $x_{j+1}=y$, because the current point changes only when its distance to $t$ decreases strictly. Otherwise, \Cref{lem:euclid-contraction} with $x=x_j$ and $\Delta=\Delta_j$, together with \Cref{thm:sampler}\ref{it:dom}, shows that each sample $\bar c$ of stage $j$ satisfies $\|x_j+2B\bar c-t\|^2\le\|y-t\|^2+\Delta_{j+1}$ with probability at least $\theta$, conditionally on the earlier samples. Hence, all $\mathcal M$ samples fail with probability at most
\[
 (1-\theta)^{\mathcal M}\ \le\ e^{-\theta\mathcal M}\ \le\ \frac{\varepsilon}{4\cdot2^NJ},
\]
where we used $\log_2z\ge\log z$ for $z\ge1$. If some sample succeeds, then $x_{j+1}$ is at least as close to $t$ as $x_j+2B\bar c$ for that sample, which proves the claim.

By a union bound over the $J$ stages, $\|x_J-t\|^2\le\|y-t\|^2+\Delta_J$ with probability at least $1-\varepsilon/(4\cdot2^N)$. 
Since $\Delta_J<2\iota(G)\le\lambda_1(2L)^2/2$, \Cref{lem:parity-geometry}\ref{it:geo-equal} then gives $x_J=y$.
\end{proof}

\Cref{lem:euclid-resources} bounds the resources of \Cref{alg:parity}. It also applies to an oracle that may err, provided that each answer is an integer vector of bounded bit length, a condition enforced by \Cref{lem:bounded-error}.

\begin{lem}\label[lem]{lem:euclid-resources}
\Cref{alg:parity} makes at most $2^NJ\mathcal M$ oracle calls, each on a full-rank sublattice of $2L$ given by a $\poly(\ell)$-bit Gram matrix. If every oracle answer is an integer vector of $\poly(\ell)$ bits, in particular if the oracle is exact, then the algorithm uses $2^{O(N)}\poly(\ell)$ other bit operations and $\poly(\ell)$ space outside the oracle.
\end{lem}
\begin{proof}
The algorithm makes $2^NJ\mathcal M$ calls of the sampler, and each of them makes at most one oracle call by \Cref{thm:sampler}\ref{it:res}. The input $(4G,4\Delta_j,4\iota(G))$ of every call of the sampler has bit length polynomial in $\ell$, so \Cref{thm:sampler}\ref{it:res} bounds the bit lengths of the queries and the cost of each call by $\poly(\ell)$. Since $\mathcal M=2^{O(N)}\poly(\ell)$ and $J=\poly(\ell)$, the number of bit operations is $2^{O(N)}\poly(\ell)$.

For the space bound, the current point $Bc$ satisfies $\|Bc-t\|^2\le U$ at all times, because it starts at a point $Ba$ with $\|Ba-t\|^2\le U$ and its distance to $t$ never increases. By~\eqref{eq:iota}, $\|c-\bar t\|^2\le U/\iota(G)$, so $c$ has $\poly(\ell)$ bits. The same holds for $c_{\rm old}$ and $c_{\rm best}$. Every sample lies in $\mathcal B(2\sqrt U)$ by \Cref{thm:sampler}\ref{it:support}, and~\eqref{eq:iota} bounds its coefficients in the same way. Besides the workspace of one call of the sampler, which is reused, the algorithm stores three coefficient vectors and the loop counters.
\end{proof}

\begin{proof}[Proof of \Cref{thm:main} for $p=2$]
Suppose first that the oracle is exact. Let $y$ be a closest vector of $L$ to $t$, and let $a\in\{0,1\}^N$ satisfy $y\in Ba+2L$. By \Cref{lem:euclid-coset}, with probability at least $1-\varepsilon$ the iteration for $a$ ends at $y$. 
In this event, the returned coefficient vector $c_{\mathrm{best}}$ represents $Bc_{\mathrm{best}}$, which is at least as close to $t$ as $y$ and hence is a closest vector.
\Cref{lem:euclid-resources} gives the bounds (i)--(iii) of \Cref{thm:main}.

To output every closest vector, we run \Cref{alg:parity} twice with independent random bits and modify the final comparison of each coset. Let $x_a^{(1)}$ denote the point at the end of the iteration for $a$ in the first run, and put $\mu=\min_a\|x_a^{(1)}-t\|^2$.
In the second run, define $c_a^{(2)}$ and $x_a^{(2)}=Bc_a^{(2)}$ analogously, and output $c_a^{(2)}$ whenever $\|x_a^{(2)}-t\|^2=\mu$.
By \Cref{lem:parity-geometry}, every coset contains at most one closest vector. By \Cref{lem:euclid-coset} and a union bound over the at most $2\cdot2^N$ iterations for cosets that contain a closest vector, with probability at least $1-\varepsilon/2$ every such iteration ends at the closest vector in both runs. In this event $\mu=\dist(t,L)^2$, and
the second run outputs the coefficient vectors representing exactly the closest vectors, each of them once.

Finally, suppose that the oracle is replaced by an SVP algorithm that succeeds with probability at least $2/3$. We run the algorithm, in either variant, with $\varepsilon/2$ in place of $\varepsilon$, so that $\mathcal M$ is computed for $\varepsilon/2$. Before its first oracle call, the algorithm computes $T_{\rm o}=2^{N+1}J\mathcal M$, which by \Cref{lem:euclid-resources} bounds the number of oracle calls of two runs, and a bound $\ell_0=\poly(\ell)$ on the bit lengths of all queries and of their shortest vectors (see the remark after \Cref{lem:bounded-error}). \Cref{lem:bounded-error} with $\varepsilon_1=\varepsilon_2=\varepsilon/2$ shows that the success probability is at least $1-\varepsilon$. Each oracle call is answered by $\lceil\log_3(2T_{\rm o}/\varepsilon)\rceil=\poly(\ell)$ runs of the SVP algorithm, and every answer that passes the check of \Cref{lem:bounded-error} has at most $\ell_0$ bits, so \Cref{lem:euclid-resources} applies.
\end{proof}

\subsection{Polynomial-Space Consequences}\label{sec:polyspace-consequences}

We first present the parameterized transfer underlying the Euclidean bounds in \Cref{cor:euclid-consequences}.

\begin{cor}\label[cor]{cor:polyspace}
Suppose that an algorithm solves exact SVP with probability at least $2/3$ on every rank-$N$ lattice given by an $\ell_0$-bit integer basis, in time $T_1(N,\ell_0)$ and space $T_2(N,\ell_0)$ on every execution, where $T_1$ and $T_2$ are nondecreasing in $\ell_0$. Then, there is a polynomial $P$ such that, for every rational $\varepsilon\in(0,1/2)$, exact CVP on rank-$N$ inputs of bit length $\ell$, where $\ell$ includes the bit length of $\varepsilon$, can be solved with probability at least $1-\varepsilon$ in time $2^{O(N)}P(\ell)\,T_1(N,P(\ell))$ and space $P(\ell)+T_2(N,P(\ell))$.
\end{cor}
\begin{proof}
Apply \Cref{thm:main} with $p=2$ and the SVP algorithm in place of the oracle, as in the detailed proof in \Cref{sec:euclid-deferred}. All queries have rank $N$ and $\poly(\ell)$ bits, and there are $2^{O(N)}\poly(\ell)$ of them. \Cref{lem:gram-integer} converts each query into an integer basis of $\poly(\ell)$ bits whose shortest coefficient vectors are shortest for the query. Using \Cref{lem:bounded-error} as in \Cref{sec:euclid-deferred}, we run the reduction with $\varepsilon/2$ in place of $\varepsilon$ and amplify the bounded-error SVP algorithm using $\poly(\ell)$ repetitions per oracle call, so that the oracle simulation fails with probability at most $\varepsilon/2$. The queries are answered one at a time, so the space used by the SVP algorithm is reused.
\end{proof}

The reduction of Goldreich, Micciancio, Safra and Seifert~\cite{GMSS99} goes in the opposite direction. Together with \Cref{cor:polyspace} it shows that the two problems have the same time complexity for randomized polynomial-space algorithms, up to a factor $2^{O(N)}\poly(\ell)$.

\begin{cor}\label[cor]{cor:equiv}
Let $f$ be a function. If exact Euclidean SVP on lattices of rank $N$ can be solved with probability at least $2/3$ in time $f(N)\poly(\ell)$ and space $\poly(\ell)$, then so can exact Euclidean CVP in time $2^{O(N)}f(N)\poly(\ell)$. Conversely, if exact Euclidean CVP on lattices of rank $N$ can be solved with probability at least $2/3$ in time $f(N)\poly(\ell)$ and space $\poly(\ell)$, then so can exact Euclidean SVP.
\end{cor}
\begin{proof}
The first statement is \Cref{cor:polyspace} with $T_1(N,\ell_0)=f(N)\poly(\ell_0)$, $T_2(N,\ell_0)=\poly(\ell_0)$ and $\varepsilon=1/3$. For the second, let $L=B\Z^N$ with columns $b_1,\ldots,b_N$. For $i=1,\ldots,N$, let $L^{(i)}\subseteq L$ be the sublattice of vectors whose $i$th coefficient is even, and let $y_i$ be a closest vector of $L^{(i)}$ to $b_i$. The vector $y_i-b_i\in L$ has an odd $i$th coefficient, so it is nonzero. Let $Bz$ be a shortest nonzero vector of $L$. Some coefficient $z_i$ is odd, since otherwise $B(z/2)$ would be a shorter nonzero vector of $L$. Then, $Bz+b_i\in L^{(i)}$, so $\|y_i-b_i\|\le\|Bz\|$, and the shortest of $y_1-b_1,\ldots,y_N-b_N$ is a shortest nonzero vector of $L$. The $N$ instances of CVP have rank $N$ and $\poly(\ell)$ bits, and they are solved one after another. 
Running the CVP algorithm independently $\lceil\log_3(3N)\rceil$ times on each instance and keeping the closest valid lattice vector makes all $N$ answers correct with probability at least $2/3$.
\end{proof}

We next justify the polynomial-space instantiation in \Cref{cor:lp-consequences}. 
For every fixed integer $p\ge3$, the reduction of \Cref{thm:main} can be instantiated with the polynomial-space SVP algorithm of Feng, Gao and Liu~\cite[Theorem~4.1]{FGL26}, using \Cref{lem:gram-integer,lem:bounded-error}.
\Cref{lem:gram-integer} converts each rank-$N$ rational Gram-matrix query into an equivalent rank-$N$ integer-basis instance of polynomial bit length, and \Cref{lem:bounded-error} simulates the exact SVP oracle using the bounded-error SVP algorithm with only polynomial overhead. Since $(Cp)^N=2^{O(N)}$, which is $N^{o(N)}$ for fixed $p$, the $N^{N/(4e)+o(N)}\poly(\ell)$ running time of~\cite[Theorem~4.1]{FGL26} absorbs the overhead of the reduction. Thus, exact $\ell_p$-CVP on full-rank lattices with arbitrary rational targets can be solved in randomized time $N^{N/(4e)+o(N)}\poly(\ell)$ and polynomial space. The oracle calls are made sequentially, so the workspace used to answer them can be reused.

\subsection{Deferred Proofs for the \texorpdfstring{$\ell_p$}{lp} Reduction}\label{sec:lp-deferred}
We first prove Lemmas~\ref{lem:even-midpoint}--\ref{lem:lp-contraction}.
We then give the full pseudocode and parameter choices for \Cref{alg:lp}, analyze its correctness and resource bounds, and complete the proof of \Cref{thm:main} for $p\ge3$.

\subsubsection{Geometric and Progress Lemmas}
\label{sec:lp-geometric-progress}

\begin{restatedstatement}{lem:even-midpoint}
Let $y$ be a minimizer of $F_t$ on $L$, let $\Delta>0$, and let $x\in y+2L$ satisfy $F_t(x)\le F_t(y)+\Delta$. Write $\eta=y-x$, $\sigma=x-t$ and $\sigma'=y-t$. Then,
\[
 \sum_i |\eta_i|^p\le2^p\Delta,
 \qquad \sum_i \max\{|\sigma_i|,|\sigma'_i|\}^{p-2}\eta_i^2\le4\Delta,
 \qquad \sum_i\bigl(|\sigma'_i|^{p/2}-|\sigma_i|^{p/2}\bigr)^2\le2\Delta.
\]
\end{restatedstatement}

\begin{proof}[Proof of \Cref{lem:even-midpoint}]
For $a,b\in\R$ let $g(a,b)=\frac12(|a|^p+|b|^p)-|\frac{a+b}2|^p$. Since $(x+y)/2\in L$ and $y$ is a minimizer,
\begin{equation}\label{eq:g-sum}
 \sum_ig(\sigma_i,\sigma'_i)\ =\ \tfrac12\bigl(F_t(x)+F_t(y)\bigr)-F_t\bigl(\tfrac{x+y}2\bigr)\  \le\ \tfrac12\bigl(F_t(x)-F_t(y)\bigr)\ \le\ \tfrac\Delta2 .
\end{equation}
We bound $g$ from below in two ways. Put $\bar a=(a+b)/2$, $\omega=|a-b|/2$ and $M=|\bar a|+\omega=\max\{|a|,|b|\}$, and assume $\omega>0$, since both bounds are trivial for $\omega=0$. Taylor's formula with integral remainder, applied to $\phi$ at $\bar a$ in the directions $\pm\omega$, gives
\[
 g(a,b)=\frac{p(p-1)}2\int_0^\omega(\omega-\alpha)\bigl(|\bar a+\alpha|^{p-2}+|\bar a-\alpha|^{p-2}\bigr)\,d\alpha .
\]
One of $|\bar a+\alpha|$ and $|\bar a-\alpha|$ equals $|\bar a|+\alpha\ge(\alpha/\omega)M$. Hence,
\[
 g(a,b)\ \ge\ \frac{p(p-1)}2\,\frac{M^{p-2}}{\omega^{p-2}}\int_0^\omega(\omega-\alpha)\alpha^{p-2}\,d\alpha\ =\ \frac{M^{p-2}\omega^2}2\ =\ \frac18M^{p-2}(a-b)^2 .
\]
For the second bound, the triangle inequality and convexity of $u\mapsto u^{p/2}$ on $[0,\infty)$ give
\[
 \Bigl|\frac{a+b}2\Bigr|^p\le\Bigl(\frac{|a|+|b|}2\Bigr)^p\le\Bigl(\frac{|a|^{p/2}+|b|^{p/2}}2\Bigr)^2,
\]
so $g(a,b)\ge\frac14(|a|^{p/2}-|b|^{p/2})^2$. Summing each bound over $i$ with $(a,b)=(\sigma_i,\sigma'_i)$ and using~\eqref{eq:g-sum} proves the second and third inequalities. The first follows from the second, because $|\eta_i|\le2\max\{|\sigma_i|,|\sigma'_i|\}$ gives $|\eta_i|^p\le2^{p-2}\max\{|\sigma_i|,|\sigma'_i|\}^{p-2}\eta_i^2$.
\end{proof}

\begin{restatedstatement}{lem:weighted}
Let $y$ be a minimizer of $F_t$ on $L$, let $\Delta>0$, and let $x\in y+2L$ satisfy $F_t(x)\le F_t(y)+\Delta$. With~\eqref{eq:even-regularizer},
\[
 \|y-x\|_{H_x}^2\le12\Delta,
 \qquad
 \sum_i\frac{\beta_i^{p-2}}{h_i}\le7N,
 \qquad
 h_i\beta_i^2\ge\frac\Delta N\quad\text{for every }i.
\]
\end{restatedstatement}

\begin{proof}[Proof of \Cref{lem:weighted}]
We use the notation $\eta,\sigma,\sigma'$ of \Cref{lem:even-midpoint}, so that $h_i=|\sigma_i|^{p-2}+\rho$ and $\beta_i=\max\{|\sigma'_i|,\beta\}$. By H\"older's inequality and \Cref{lem:even-midpoint}, $\|\eta\|^2\le N^{1-2/p}(\sum_i|\eta_i|^p)^{2/p}\le4N^{1-2/p}\Delta^{2/p}$. With $\rho<2(\Delta/N)^{1-2/p}$ this gives $\rho\|\eta\|^2<8\Delta$. Since $|\sigma_i|\le\max\{|\sigma_i|,|\sigma'_i|\}$, \Cref{lem:even-midpoint} also gives $\sum_i|\sigma_i|^{p-2}\eta_i^2\le4\Delta$. Adding the two bounds proves $\|\eta\|_{H_x}^2\le12\Delta$.

Since $|\sigma'_i|^{p/2}\le|\sigma_i|^{p/2}+\bigl||\sigma'_i|^{p/2}-|\sigma_i|^{p/2}\bigr|$, raising to the power $2-4/p\in(0,2)$ and using $(a+b)^{2-4/p}\le2\bigl(a^{2-4/p}+b^{2-4/p}\bigr)$ for $a,b\ge0$ gives
\[
 |\sigma'_i|^{p-2}\ \le\ 2|\sigma_i|^{p-2}+2\,\bigl||\sigma'_i|^{p/2}-|\sigma_i|^{p/2}\bigr|^{2-4/p}.
\]
Divide by $h_i$, which is at least $|\sigma_i|^{p-2}$ and at least $\rho$, and bound the sum of the last terms by H\"older's inequality and the third bound of \Cref{lem:even-midpoint}:
\[
 \sum_i\frac{|\sigma'_i|^{p-2}}{h_i}\ \le\ 2N+\frac2\rho\,N^{2/p}(2\Delta)^{1-2/p}\ \le\ 2N+2^{2-2/p}N\ \le\ 6N,
\]
where the middle step uses $\rho\ge(\Delta/N)^{1-2/p}$. Since $\beta_i^{p-2}\le|\sigma'_i|^{p-2}+\beta^{p-2}$ and $\beta^{p-2}\le\rho\le h_i$, the second claim follows. Finally $h_i\beta_i^2\ge\rho\beta^2\ge\beta^p=\Delta/N$.
\end{proof}

\begin{restatedstatement}{lem:taylor}
For $a,w\in\R$ and $\beta_0>0$ with $\beta_0\ge|a|$,
\[
 0\ \le\ \phi(a+w)-\phi(a)-\phi'(a)w
 \ \le\ \tfrac{p^2}{2}\beta_0^{p-2}e^{p|w|/\beta_0}w^2 .
\]
\end{restatedstatement}

\begin{proof}[Proof of \Cref{lem:taylor}]
The lower bound holds because $\phi$ is convex. By Taylor's theorem the middle term equals $\frac12\phi''(\zeta)w^2$ for some $\zeta$ between $a$ and $a+w$. As $|\zeta|\le \beta_0+|w|$,
\[
 \phi''(\zeta)=p(p-1)|\zeta|^{p-2}\le p^2\beta_0^{p-2}(1+|w|/\beta_0)^{p-2}\le p^2\beta_0^{p-2}e^{p|w|/\beta_0}.\qedhere
\]
\end{proof}

\begin{restatedstatement}{lem:lp-contraction}
Let $y$ be a minimizer of $F_t$ on $L$, let $\Delta>0$, and let $x\in y+2L$ satisfy $x\ne y$ and $F_t(x)\le F_t(y)+\Delta$. Let $H_x$ be defined by~\eqref{eq:even-regularizer}, let $\mathcal D$ be a distribution on $2L$ satisfying~\eqref{eq:dom} with $K=2L$, $H=H_x$, $r=4\sqrt\Delta$ and some $\gamma\in(0,1]$, and let $V\sim\mathcal D$. Then
\[
 \Pr\bigl[F_t(x+V)\le F_t(y)+\Delta/4\bigr]
 \ \ge\ \tfrac14\cdot(188p)^{-N}\gamma.
\]  
\end{restatedstatement}

\begin{proof}[Full proof of \Cref{lem:lp-contraction}]
We use the notation $\eta,\sigma,\sigma'$ of \Cref{lem:even-midpoint} and $h_i,\beta_i$ of~\eqref{eq:even-regularizer}, and we let $K=2L$. We construct a set $E\subseteq K\cap \mathcal B_{H_x}(4\sqrt\Delta)$ such that every $z\in E$ satisfies $F_t(x+z)-F_t(y)<\Delta/4$ and $|E\setminus\{0\}|\ge\frac14(188p)^{-N}|K\cap \mathcal B_{H_x}(4\sqrt\Delta)|$. Then~\eqref{eq:dom} gives the lemma. 
The set is $E=\eta+S_-$, where $S_-$ will be chosen so that every $w\in S_-$ satisfies $F_t(y+w)-F_t(y)<\Delta/4$. We define $S_-$ through a box and a sign condition.
The box gives the bound on each $|w_i|$ required by \Cref{lem:taylor} at a smaller cost in the count than a ball, and the sign condition ensures that the linear term of the Taylor expansion of $F_t$ at $y$ is nonpositive.

Let $\tau=(\Delta/(32N))^{1/2}/p$, and define
\[
 Q=\{w\in\R^N:\ h_iw_i^2\le\tau^2\ \text{for all }i\},\qquad
 S_-=\Bigl\{w\in K\cap Q:\ \sum_i\phi'(\sigma'_i)\,w_i\le0\Bigr\}.
\]
The set $K\cap Q$ is symmetric and the sum in the definition of $S_-$ changes sign with $w$, so $w$ or $-w$ lies in $S_-$ for every $w\in K\cap Q$. Hence $|S_-|\ge|K\cap Q|/2$.

\begin{claim}[Improvement]\label[claim]{cl:box-improves}
For every $w\in S_-$, $F_t(y+w)-F_t(y)<\Delta/4$.
\end{claim}
\begin{proof}
Let $w\in S_-$. By \Cref{lem:weighted}, $h_i\beta_i^2\ge\Delta/N$, so
\[
 \frac{p|w_i|}{\beta_i}\ \le\ \frac{p\tau}{\sqrt{h_i}\,\beta_i}\ \le\ \frac1{\sqrt{32}},
 \qquad
 p^2w_i^2\ \le\ \frac{p^2\tau^2}{h_i}\ =\ \frac{\Delta}{32N h_i}.
\]
We have
\[
 F_t(y+w)-F_t(y)\ =\ \sum_i\phi'(\sigma'_i)\,w_i+\sum_i\bigl(\phi(\sigma'_i+w_i)-\phi(\sigma'_i)-\phi'(\sigma'_i)w_i\bigr),
\]
and the first sum is at most $0$ because $w\in S_-$. Bounding each term of the second sum by \Cref{lem:taylor} with $a=\sigma'_i$ and $\beta_0=\beta_i\ge|\sigma'_i|$, and then using \Cref{lem:weighted}, we obtain
\[
 F_t(y+w)-F_t(y)\ \le\ \frac{e^{1/\sqrt{32}}}2\sum_ip^2\beta_i^{p-2}w_i^2\ \le\ \frac{e^{1/\sqrt{32}}}2\cdot\frac{\Delta}{32N}\sum_i\frac{\beta_i^{p-2}}{h_i}\ \le\ \frac{7}{64}e^{1/\sqrt{32}}\,\Delta .
\]
Since $\frac{7}{64}e^{1/\sqrt{32}}<\frac14$, the right-hand side is less than $\Delta/4$.
\end{proof}

\begin{claim}[Containment in the sampling ball]\label[claim]{cl:box-contained}
$\eta+S_-\subseteq K\cap \mathcal B_{H_x}(4\sqrt\Delta)$.
\end{claim}
\begin{proof}
For $w\in Q$, $\|w\|_{H_x}\le\sqrt N\tau=\sqrt{\Delta/32}/p<0.06\sqrt\Delta$ since $p\ge3$. With \Cref{lem:weighted}, $\|\eta+w\|_{H_x}\le(\sqrt{12}+0.06)\sqrt\Delta<4\sqrt\Delta$.
\end{proof}

\begin{claim}[Counting bound]\label[claim]{cl:box-count}
$|K\cap \mathcal B_{H_x}(4\sqrt\Delta)|\le(188p)^N\,|K\cap Q|$.
\end{claim}
\begin{proof}
The linear map $w\mapsto H_x^{1/2}w$ sends $K$ to the lattice $H_x^{1/2}K$, the set $Q$ to the cube $[-\tau,\tau]^N$, and $\mathcal B_{H_x}(4\sqrt\Delta)$ to $\mathcal B(4\sqrt\Delta)$. Since $\tau\le4\sqrt\Delta/\sqrt N$, \Cref{lem:cover}(ii) with $\alpha=\tau$ and~\eqref{eq:vn} cover $\mathcal B(4\sqrt\Delta)$ by at most
\[
 \omega_N\cdot\Bigl(\frac{8\sqrt\Delta}{\tau}\Bigr)^N\ \le\ \Bigl(\frac{2\pi e}N\Bigr)^{N/2}\bigl(8\sqrt{32N}\,p\bigr)^N\ =\ \bigl(64\sqrt{\pi e}\,p\bigr)^N\ <\ (188p)^N
\]
translates of $[0,\tau]^N$, which is a translate of $\frac12[-\tau,\tau]^N$. \Cref{lem:count} applied to $H_x^{1/2}K$ proves the claim.
\end{proof}

Let $E=\eta+S_-$. By \Cref{cl:box-improves}, every $z\in E$ satisfies $F_t(x+z)-F_t(y)<\Delta/4$. Since $|E|=|S_-|$ and $E$ contains $\eta\ne0$, the argument in the proof of \Cref{lem:euclid-contraction} gives $|E\setminus\{0\}|\ge|S_-|/2$. By \Cref{cl:box-contained,cl:box-count} and~\eqref{eq:dom},
\[
 \Pr[V\in E]=\mathcal D(E)\ \ge\ \gamma\,\frac{|S_-|/2}{(188p)^N|K\cap Q|}\ \ge\ \gamma\,\frac{|S_-|/2}{2(188p)^N|S_-|}\ =\ \tfrac14\,\gamma\,(188p)^{-N}.\qedhere
\]
\end{proof}

\subsubsection{Full pseudocode and parameters}

\Cref{alg:lp} presents our algorithm for exact $\ell_p$-CVP for integer $p\ge3$, following the structure of \Cref{alg:parity}.
It represents a lattice point $Bc$ by its coefficient vector $c\in\Z^N$ and evaluates $F_t(Bc)$ exactly in rational arithmetic.

\paragraph{Description.}
The algorithm processes the cosets $Ba+2L$, $a\in\{0,1\}^N$, one at a time and runs the same fixed sequence of stages in each coset. At the start of stage $j$, it stores the current point $x_{\rm old}=Bc_{\rm old}$ and computes the diagonal metric $H=H_{x_{\rm old}}$ from~\eqref{eq:even-regularizer} with $\Delta=\Delta_j$. This metric is kept fixed throughout the stage. Every sampled correction is added to the same point $x_{\rm old}$, and the algorithm keeps the best point found so far according to $F_t$. Thus, the structure is the same as in \Cref{alg:parity}, except that the metric is recomputed at the start of each stage and the sampler is run in that weighted metric.

\paragraph{Parameters.}
Let $d$ be the least common multiple of the denominators of the entries of $B$ and $t$. The algorithm uses the numbers
\[
 U=\sum_i\Bigl(|t_i|+\sum_k|B_{ik}|\Bigr)^p,\qquad \Delta_j=U/4^j\ \ (j\ge0),\qquad J=\min\{j\ge0:\ \Delta_j<d^{-p}\}.
\]
For $a\in\{0,1\}^N$, we have $|(Ba)_i-t_i|\le|t_i|+\sum_k|B_{ik}|$, so $F_t(Ba)\le U$. The number $J$ of stages is chosen as follows. Every coordinate of a lattice point or of $t$ lies in $d^{-1}\Z$, so every value of $F_t$ on $L$ lies in $d^{-p}\Z$. Hence, a lattice point $x$ with $F_t(x)\le F_t(y)+\Delta_J$ for a minimizer $y$ satisfies $F_t(x)=F_t(y)$. Since $U$ and $d^{-p}$ are positive rationals of $\poly(p,\ell)$ bit length, we have $J=O(1+\log_2\max\{1,Ud^p\})=\poly(p,\ell)$. Each row of the nonsingular matrix $B$ has a nonzero entry, whose absolute value is at least $1/d$. Hence $U\ge Nd^{-p}$, and in particular $J\ge1$.

At stage $j$, the lattice $2L$ in the metric $H=H_{x_{\rm old}}$ has Gram matrix $(2B)^TH(2B)$. 
A nonzero vector of $2L$ has a coordinate of absolute value at least $2/d$, so its squared $H$-length is at least $4\rho/d^2$. The algorithm calls the sampler of \Cref{thm:sampler} on $((2B)^TH(2B),16\Delta_j,4\rho/d^2)$, that is, with radius $r=4\sqrt{\Delta_j}$ and $\lambda_0^2=4\rho/d^2$.

To bound the success probability uniformly over the stages, we bound $r/\lambda_0$. For $j<J$, we have $\Delta_j\ge d^{-p}$, and since $0<1-2/p<1$,
\[
 \rho\ \ge\ (\Delta_j/N)^{1-2/p}\ \ge\ \min\{1,\Delta_j/N\}\ \ge\ d^{-p}/N .
\]
Hence $(r/\lambda_0)^2=16\Delta_jd^2/(4\rho)\le4Ud^{p+2}N$, and $4Ud^{p+2}N\ge1$ because $U\ge Nd^{-p}$. Thus $\log_2\max\{1,r/\lambda_0\}\le\lceil\log_2(4Ud^{p+2}N)\rceil$ in every stage. By \Cref{lem:lp-contraction} and \Cref{thm:sampler}\ref{it:dom}, a sample of stage $j$ therefore succeeds with probability at least
\[
 \theta=\frac{(1692p)^{-N}}{2048\,N\,(\lceil\log_2(4Ud^{p+2}N)\rceil+4)^2},
 \qquad\text{and we put}\qquad
 \mathcal M=\theta^{-1}\Bigl\lceil\log_2\frac{4\cdot2^NJ}{\varepsilon}\Bigr\rceil .
\]
Both $\theta^{-1}$ and $\mathcal M$ are integers, and $\mathcal M=(1692p)^N\poly(p,\ell)$.

\begin{algorithm}[t]
\DontPrintSemicolon
\Input{Nonsingular $B\in\Q^{N\times N}$, target $t\in\Q^N$, integer $p\ge3$, rational $\varepsilon\in(0,1/2)$}
\Output{Coefficient vector of a closest lattice point in the $\ell_p$ norm}
\vspace{2mm}
$c_{\rm best}\gets0$\;
\For{$a\in\{0,1\}^N$}{
 $c\gets a$\;
 \For{$j=0,\ldots,J-1$}{
  $c_{\rm old}\gets c$\;
  $H\gets H_{Bc_{\rm old}}$ computed by~\eqref{eq:even-regularizer} with $\Delta=\Delta_j$\;
  \For{$\mathcal M$ times}{
   $\bar c\gets$ output of \Cref{alg:sampler} on $\bigl((2B)^TH(2B),\,16\Delta_j,\,4\rho/d^2\bigr)$\;
   \lIf{$F_t(B(c_{\rm old}+2\bar c))<F_t(Bc)$}{$c\gets c_{\rm old}+2\bar c$}
  }
 }
 \lIf{$F_t(Bc)<F_t(Bc_{\rm best})$}{$c_{\rm best}\gets c$}
}
\Return{$c_{\rm best}$}\;
\caption{Exact CVP in the $\ell_p$ norm for an integer $p\ge3$}
\label{alg:lp}
\end{algorithm}

\subsubsection{Correctness and Resource Analysis}
We prove the correctness and resource bounds of \Cref{alg:lp}, completing the proof of \Cref{thm:main} for $p\ge3$.

\begin{lem}\label[lem]{lem:lp-coset}
Assume that the oracle is exact. Let $y$ be a minimizer of $F_t$ on $L$, and let $a\in\{0,1\}^N$ satisfy $y\in Ba+2L$. Then, with probability at least $1-\varepsilon/(4\cdot2^N)$, the iteration of \Cref{alg:lp} for $a$ ends with $Bc=y$.
\end{lem}
\begin{proof}
The proof follows that of \Cref{lem:euclid-coset}. Let $x_j$ be the current point at the start of stage $j$, and let $x_J$ be the point at the end of the last stage. All these points lie in $y+2L$, and
\[
 F_t(x_0)\le U=\Delta_0\le F_t(y)+\Delta_0,
\]
so the invariant holds initially. Fix $j<J$ and condition on the computation up to the start of stage $j$, assuming that $F_t(x_j)\le F_t(y)+\Delta_j$. If $x_j=y$, then $x_{j+1}=y$, because the current point changes only when $F_t$ decreases strictly. Otherwise, \Cref{lem:lp-contraction} with $x=x_j$ and $\Delta=\Delta_j$, together with \Cref{thm:sampler}\ref{it:dom}, shows that each sample of stage $j$ yields a point $x'$ with $F_t(x')\le F_t(y)+\Delta_{j+1}$ with probability at least $\theta$, conditionally on the earlier samples. As in the proof of \Cref{lem:euclid-coset}, all $\mathcal M$ samples fail with probability at most $\varepsilon/(4\cdot2^NJ)$. A union bound over the $J$ stages shows that $F_t(x_J)\le F_t(y)+\Delta_J$ with probability at least $1-\varepsilon/(4\cdot2^N)$.
In this event, $F_t(x_J)=F_t(y)$ by the choice of $J$, so $x_J$ is also a minimizer of $F_t$ on $L$.
Since $F_t$ is strictly convex, two distinct minimizers $y,y'$ in the same coset of $2L$ would give a lattice point $(y+y')/2$ with $F_t((y+y')/2)<F_t(y)$. Hence, $x_J=y$.
\end{proof}

\begin{lem}\label[lem]{lem:lp-resources}
\Cref{alg:lp} makes at most $2^NJ\mathcal M$ oracle calls, each on a full-rank sublattice of $2L$ in a diagonal metric, given by a $\poly(p,\ell)$-bit Gram matrix. If every oracle answer is an integer vector of $\poly(p,\ell)$ bits, in particular if the oracle is exact, then the algorithm uses $(Cp)^N\poly(p,\ell)$ other bit operations, for an absolute constant $C$, and $\poly(p,\ell)$ space outside the oracle.
\end{lem}
\begin{proof}
Each of the $2^NJ$ stages makes $\mathcal M$ calls of the sampler, and each of them makes at most one oracle call by \Cref{thm:sampler}\ref{it:res}. The current point satisfies $F_t(Bc)\le U$ at all times, because it starts at a point $Ba$ with $F_t(Ba)\le U$ and $F_t$ never increases. Hence, $|(Bc)_i-t_i|\le\max\{1,U\}$, and since $c=B^{-1}(Bc)$, the vector $c$ has $\poly(p,\ell)$ bits. The same holds for $c_{\rm old}$ and $c_{\rm best}$. 
The number \(\rho\) lies between \(d^{-p}/N\) and \(2\max\{1,U\}\), so \(H\), \(4\rho/d^2\), and \(16\Delta_j\) have \(\poly(p,\ell)\) bits. Hence, \Cref{thm:sampler}\ref{it:res} implies that every SVP query has \(\poly(p,\ell)\)-bit Gram matrix. For an exact oracle, every shortest coefficient vector for such a query has \(\poly(p,\ell)\) bits by the coefficient bound following \Cref{lem:bounded-error}. 
Under the stated assumption on oracle answers, \Cref{thm:sampler}\ref{it:res} therefore implies that each call of the sampler uses $\poly(p,\ell)$ non-oracle bit operations and space.
Every sample lies in $\mathcal B_H(4\sqrt U)$ by \Cref{thm:sampler}\ref{it:support}. Since $H\succeq\rho I$, its ambient coordinates have $\poly(p,\ell)$ bits; applying the rational matrix $(2B)^{-1}$ then shows that its coefficient vector with respect to the basis $2B$ also has $\poly(p,\ell)$ bits. Since $\mathcal M=(1692p)^N\poly(p,\ell)$ and $J=\poly(p,\ell)$, the number of bit operations is $(Cp)^N\poly(p,\ell)$. Besides the workspace of one call of the sampler, the algorithm stores three coefficient vectors, one diagonal matrix, and the loop counters.
\end{proof}

\begin{proof}[Proof of \Cref{thm:main} for $p\ge3$]
The proof follows that for $p=2$, with \Cref{lem:lp-coset,lem:lp-resources} in place of \Cref{lem:euclid-coset,lem:euclid-resources}. 
By \Cref{lem:lp-coset}, the coefficient vector returned by \Cref{alg:lp} represents a minimizer of $F_t$ on $L$, and hence an $\ell_p$-closest vector, with probability at least $1-\varepsilon$.
Each coset of $2L$ contains at most one minimizer by the strict convexity argument at the end of the proof of \Cref{lem:lp-coset}. Applying the two-run procedure of \Cref{sec:euclid-deferred} to \Cref{alg:lp}, with comparisons made using $F_t$, outputs every minimizer exactly once with probability at least $1-\varepsilon/2$. 
For an SVP algorithm that succeeds with probability at least $2/3$, \Cref{lem:bounded-error} applies as in the proof for $p=2$, with $\varepsilon/2$ in place of $\varepsilon$ and $T_{\rm o}=2^{N+1}J\mathcal M$.
\end{proof}

%% file: 8_conclusion.tex
\section{Conclusion}\label{sec:conclusion}
We have presented a randomized, rank-preserving reduction from exact $\ell_p$-CVP to exact Euclidean SVP for every integer $p\ge2$, using only polynomial working space outside the oracle.
For $p=2$, our reduction shows that the randomized polynomial-space time complexities of exact Euclidean SVP and CVP agree up to a $2^{O(N)}\poly(\ell)$ factor. For every fixed integer $p\ge3$, it yields exact $\ell_p$-CVP algorithms for arbitrary rational targets with either single-exponential time and space or $N^{N/(4e)+o(N)}\poly(\ell)$ time and polynomial space. The reduction combines local search within cosets of $2L$, counting arguments for improving corrections, and SVP-based sampling. An interesting direction for future work is to reduce the exponential overhead while preserving its rank and space guarantees.

%% file: a_oracle_details.tex
\section{Oracle Details}\label[appendix]{app:implementation}

This appendix presents two implementation facts used in the reductions. First, we show that an SVP query given by a positive definite rational Gram matrix can be converted, without increasing the rank, into an equivalent query on an integer basis. Second, we show how an exact SVP oracle can be replaced by a bounded-error SVP algorithm with only polynomial overhead.

\begin{lem}[Rational Gram matrices]\label[lem]{lem:gram-integer}
Exact SVP for a positive definite rational $N\times N$ Gram matrix $G$ reduces in polynomial time to exact SVP on an integer square basis of rank $N$, in the sense that every shortest coefficient vector for the new basis is a shortest coefficient vector for $G$.
\end{lem}
\begin{proof}
Let $D_0$ clear the denominators of $G$, and put $a_{\max}=\max_iG_{ii}$, $\iota=\iota(G)$ and $b_0^2=4a_{\max}/\iota$. Distinct values of $a^TGa$, $a\in\Z^N$, differ by at least $1/D_0$.
Choose a positive rational $\varepsilon_0$ satisfying
\[
  \varepsilon_0<\min\{\iota/4,\,1/(8D_0b_0^2)\}.
\]
A rational $LDL^T$ decomposition of $G$, followed by sufficiently accurate dyadic approximations of the square roots of the diagonal entries in the decomposition, gives a rational square matrix $\widetilde B$ with $\|\widetilde B^T\widetilde B-G\|_{\mathrm{op}}<\varepsilon_0$.
Here, an entrywise error bound multiplied by $N$ bounds the operator norm. Write $\widetilde G=\widetilde B^T\widetilde B$.

Since $\iota\le\lambda_{\min}(G)\le a_{\max}$, we have $\varepsilon_0<a_{\max}/4$, and the least eigenvalue of $\widetilde G$ is at least $\frac34\iota$. Let $a^*$ and $\tilde a$ be shortest coefficient vectors for $G$ and for $\widetilde G$, respectively. Comparing with unit vectors gives $a^{*T}Ga^*\le a_{\max}$ and $\tilde a^T\widetilde G\tilde a\le a_{\max}+\varepsilon_0<\frac54a_{\max}$. Hence, $\|a^*\|^2\le a_{\max}/\iota<b_0^2$ by~\eqref{eq:iota}, and $\|\tilde a\|^2<\frac54a_{\max}/(\frac34\iota)<b_0^2$. For both vectors the two quadratic forms differ by less than $\varepsilon_0b_0^2<1/(8D_0)$.
If $\tilde a^TG\tilde a>a^{*T}Ga^*$, then $\tilde a^TG\tilde a\ge a^{*T}Ga^*+1/D_0$, and therefore $\tilde a^T\widetilde G\tilde a>a^{*T}\widetilde Ga^*+3/(4D_0)$, which contradicts the choice of $\tilde a$. Hence $\tilde a$ is shortest for $G$. 

Let $D$ be a positive integer clearing the denominators of $\widetilde B$. Then $\widehat B=D\widetilde B$ is an integer basis with Gram matrix $\widehat B^T\widehat B=D^2\widetilde G$, so its shortest coefficient vectors are exactly those of $\widetilde G$. The entries of the rational $LDL^T$ decomposition are ratios of minors of $G$, and $\log(1/\varepsilon_0)$ is polynomial in the bit length of $G$ because $D_0$, $a_{\max}$ and $\iota$ are. Thus, $\varepsilon_0$ can be chosen with polynomial bit length, and all numbers involved have polynomial bit length.
\end{proof}

\begin{lem}[Bounded-error SVP algorithms]\label[lem]{lem:bounded-error}
Let $0\le\varepsilon_1<1$ and $0<\varepsilon_2<1$. Let $\mathcal R$ be a randomized algorithm with an exact SVP oracle that succeeds with probability at least $1-\varepsilon_1$ for every rule by which the oracle chooses among shortest vectors. Suppose that before its first call $\mathcal R$ computes $T_{\rm o}\ge1$ and $\ell_0$ such that, for all answers that pass the check below, it makes at most $T_{\rm o}$ calls, each on a lattice whose Gram matrix and shortest nonzero coefficient vectors have at most $\ell_0$ bits. Let $\mathcal S$ be an SVP algorithm that always halts within its time bound and returns a shortest nonzero vector with probability at least $2/3$. Answer each call by running $\mathcal S$ independently $k=\lceil\log_3(T_{\rm o}/\varepsilon_2)\rceil$ times and returning the shortest output that is a nonzero integer coefficient vector of the correct dimension and of at most $\ell_0$ bits, or aborting if there is none. Then, $\mathcal R$ succeeds with probability at least $1-\varepsilon_1-\varepsilon_2$, and each call costs $k$ runs of $\mathcal S$.
\end{lem}
\begin{proof}
The idea is to amplify each oracle call enough that, with high probability, at least one of the repeated runs returns an actual shortest vector. Conditioned on the history, all $k$ runs fail to return a shortest vector with probability at most $3^{-k}\le\varepsilon_2/T_{\rm o}$. If at least one run succeeds, then one retained output is a shortest vector; since every retained nonzero integer vector represents a lattice vector, the shortest retained output is also shortest.

Couple the execution with one that uses an exact oracle returning the same answer whenever the simulation returns a shortest vector. The two executions agree until the first call at which the simulation fails to return a shortest vector, including the case that it aborts, and a union bound over at most $T_{\rm o}$ calls bounds the probability of such a failure by $\varepsilon_2$. Therefore, the overall success probability is at least $1-\varepsilon_1-\varepsilon_2$.
\end{proof}

The coefficient bound required in \Cref{lem:bounded-error} can be computed directly from the query Gram matrix.
For a query with Gram matrix $G'$, a shortest vector with coefficient vector $a$ satisfies
\[
  \iota(G')\|a\|^2\le a^TG'a\le\min_jG'_{jj},
\]
where the upper bound follows by comparing the shortest vector with the unit coefficient vectors. Hence,
\[
  \|a\|^2\le\frac{\min_jG'_{jj}}{\iota(G')}.
\]
Thus, when $G'$ has polynomial bit length, the right-hand side has polynomial bit length, and the coordinates of every shortest coefficient vector have polynomial bit length as well.

%% file: ai_methodology.tex
\section*{AI Methodology}
We used OpenAI Codex, ChatGPT and Claude to assist with developing and checking proofs, comparing the literature, and editing. 
The tools materially affected Sections \Cref{sec:sampler-statement,sec:euclidean,sec:even,sec:sampler}, in particular the counting arguments of \Cref{lem:euclid-contraction,lem:lp-contraction}, and the short-vector sampler in \Cref{sec:sampler-statement} and its analysis in \Cref{sec:sampler}.